\documentclass[sigplan]{acmart}
\usepackage{booktabs} % For formal tables
\usepackage{todonotes}
\usepackage{paralist}
\usepackage{subcaption}

\usepackage{algpseudocode}
\usepackage{algorithm}
\usepackage{multicol}
\usepackage{multirow}
\usepackage{amsfonts}

\algnewcommand{\IIf}[1]{\State\algorithmicif\ #1\ \algorithmicthen}
\algnewcommand{\Spin}[1]{\State\algorithmicrepeat\ \algorithmicuntil \ #1}
\algnewcommand{\EndIIf}{\unskip\ \algorithmicend\ \algorithmicif}
\algnewcommand{\algorithmicgoto}{\textbf{go to}}%
\algnewcommand{\Goto}[1]{\State\algorithmicgoto~\ref{#1}}%
\algnewcommand{\IGoto}[1]{\algorithmicgoto~\ref{#1}}%
\newcommand{\lt}{<}
\newcommand{\gt}{>}

\copyrightyear{2019} 
\acmYear{2019} 
\setcopyright{acmcopyright}
\acmConference[PPoPP '19]{24th ACM SIGPLAN Symposium on Principles and Practice of Parallel Programming}{February 16--20, 2019}{Washington, DC, USA}
\acmBooktitle{24th ACM SIGPLAN Symposium on Principles and Practice of Parallel Programming (PPoPP '19), February 16--20, 2019, Washington, DC, USA}
\acmPrice{15.00}
\acmDOI{10.1145/3293883.3295730}
\acmISBN{978-1-4503-6225-2/19/02}

\acmSubmissionID{ppopp19-p175}

\begin{document}
\title{Processing Transactions in a Predefined Order}
%\titlenote{Produces the permission block, and copyright information}
%\subtitle{Extended Abstract}
%\subtitlenote{The full version of the author's guide is available as \texttt{acmart.pdf} document}

\author{Mohamed M. Saad}
\affiliation{
  \institution{Alexandria University}            %% \institution is required
  \city{Alexandria}
  \country{Egypt}                    %% \country is recommended
}
\email{msaad@alexu.edu.eg}          %% \email is recommended

\author{Masoomeh Javidi Kishi}
\affiliation{
  \institution{Lehigh University}
  \city{Bethlehem}
  \state{PA}
  \country{USA}
}
\email{maj717@lehigh.edu}          %% \email is recommended

\author{Shihao Jing}
\affiliation{
  \institution{Lehigh University}
  \city{Bethlehem}
  \state{PA}
  \country{USA}
}
\email{shj316@lehigh.edu}          %% \email is recommended

\author{Sandeep Hans}
\affiliation{
  \institution{IBM India Research Lab}            %% \institution is required
  \city{New Delhi}
  \state{Delhi}
  \country{India}                    %% \country is recommended
}
\email{shans001@in.ibm.com}          %% \email is recommended

\author{Roberto Palmieri}
\affiliation{%
  \institution{Lehigh University}
  \city{Bethlehem}
  \state{PA}
  \country{USA}
}
\email{palmieri@lehigh.edu}

\renewcommand{\shortauthors}{M.M. Saad et al.}

\begin{abstract}
In this paper we provide a high performance solution to the problem of committing transactions while enforcing a predefined order. We provide the design and implementation of three algorithms, which deploy a specialized cooperative transaction execution model. This model permits the propagation of written values along the chain of ordered transactions. We show that, even in the presence of data conflicts, the proposed algorithms are able to outperform single-threaded execution, and other baseline and specialized state-of-the-art competitors (e.g., STMLite). The maximum speedup achieved in micro benchmarks, STAMP, PARSEC and SPEC200 applications is in the range of 4.3x -- 16.5x.
\end{abstract}




\maketitle

\section{Introduction}
\label{sec:intro}

Transaction ordering intuitively means considering not just the set of transactions as input of the problem, but also the specific commit order that must be enforced for them. 
Such a formulation inherently includes a fundamental trade off between the level of parallelism achievable, given the need of committing in-order, and the performance of the single threaded execution without any software instrumentation.

Ordering tasks before their execution is a problem, mostly relevant to contexts where producing executions equivalent to a predefined order is needed in order to satisfy certain properties (e.g., a program semantics equivalent to serial execution). Examples of these deployments include: speculative loop parallelization~\cite{streit2013sambamba,gonzalez2014effective,von2008modeling,DBLP:conf/hotpar/SaadMR12,DBLP:conf/systor/SaadPR18}, and distributed computation using the state machine approach~\cite{Schneider:1990:IFS:98163.98167,DBLP:conf/osdi/Kapritsos0QCAD12,DBLP:conf/middleware/HirvePR14}.

In the former, loops designed to run sequentially are parallelized by executing their iterations concurrently and guarding memory transactions (e.g., by using Transactional Memory~\cite{herlihy93transactional} as done in~\cite{gonzalez2014effective,DBLP:conf/hotpar/SaadMR12,DBLP:conf/systor/SaadPR18}). In that case, providing an order matching the sequential one is fundamental to enforce equivalent semantics for both the parallel and sequential code. Regarding the latter, many distributed systems order tasks before executing them to guarantee that a single state machine abstraction always evolves consistently on distinct nodes. A common example of this methodology is when consensus (e.g., Paxos~\cite{DBLP:journals/tocs/Lamport98})  is employed to establish a common order among commands (or transactions) manipulating a single replicated state.

In this paper we focus on Transactional Memory (TM) as a technology to support speculative execution of tasks, and we present three algorithms to process transactions in parallel while enforcing a predefined order: \textit{Ordered Write Back} (OWB), \textit{Ordered Undo Logging} (OUL), and a lock-steal variant of OUL (OUL-Steal). These algorithms are based on two widely used techniques to merge transaction modifications into the shared state: write-back (in OWB) and write-through (in OUL and OUL-Steal).

All our implementations deploy a common design that uses a cooperative model, where transactions exchange both data and locks to increase concurrency while preserving the predefined commit order. OWB uses data forwarding for transactions that finish their execution successfully, but are not committed yet, and OUL leverages encounter time locking with the ability to pass the lock ownership to other transactions. Our cooperative model is similar to the dependency aware transactions model (DATM)~\cite{DBLP:conf/micro/RamadanRW08,ramadan2009committing}. However, DATM is not designed (and thus cannot be optimized) for committing transactions in a predefined order since it tracks all dependencies among transactions and analyzes them at run-time seeking for some correct serialization order instead of the predefined one.

We implement OWB, OUL, OUL-Steal, the ordered version of four existing well-known TM designs (i.e., TL2~\cite{Dice06transactionallocking}, NOrec~\cite{dalessandro2010norec}, and UndoLog~\cite{felber2008dynamic} with and without visible readers), and STMLite~\cite{pldi09}, a specialized solution that commits transactions in a predefined order. We conduct a performance evaluation using a set of micro-benchmarks, STAMP~\cite{caominh:stamp:iiswc:2008}, and some applications from the PARSEC and SPEC2000 benchmarks.
For determining the transaction order, we use either the index of the application main for-loop that generates the parallel code or an artificial atomic integer that we inserted as transaction order. Results have been compared agains the sequential execution of the benchmarks, as well as against their parallel execution, as provided by the original version of the applications.

Results show interesting trends: our OUL outperforms other ordered competitors consistently. In particular, the maximum speedup achieved is 4$\times$ over Ordered TL2, 4.3$\times$ over Ordered NORec, 8$\times$ over Ordered UndoLog visible, 10$\times$ over Ordered UndoLog invisible, and 5.7$\times$ over STMLite.
Interestingly, the peak gain over the sequential non-instrumented execution in micro benchmarks is 10$\times$, 16.5$\times$ in STAMP, more than 10$\times$ in PARSEC, and 30\% in SPEC2000. Also, there are configurations where the sequential execution of the applications outperforms all ordered competitors, except ours.

OWB ensures TMS1~\cite{DohertyGLM2013}, a weaker consistency condition than opacity~\cite{gue08,GuerraouiK2011book}, the most popular correctness level for TM.
TMS1 has been proved to be sufficient to guarantee safety in our model~\cite{AttiyaGHR2014}, as is the case with opacity. 
OUL achieves higher concurrency and therefore higher performance, at the cost of weakening the correctness level by ensuring Strict Serializability~\cite{bernstein1987concurrency}.

Finally, it is worth mentioning that OWB, OUL, and OUL-Steal are TM implementations meant to be integrated into runtime systems to support their speculative execution. An example of such a system is Lerna~\cite{DBLP:conf/systor/SaadPR18}. In those systems, a sandboxing~\cite{Dalessandro:2012:STM:2370816.2370843} mechanism prevents computation exceptions to be propagated outside the concurrency control engine.

%Such a TM is released as an open-source project at:\\ \url{https://bitbucket.org/mohamed-m-saad/ordertm}.
%Our implementation is released as an open-source project at \url{https://bitbucket.org/mohamed-m-saad/ordertm}.

\section{Related Work}

Transactional Memory (TM) has emerged as a technique for protecting speculative code~\cite{herlihy93transactional,pldi09}, and extracting parallelization from sequential code~\cite{streit2013sambamba,gonzalez2014effective,von2008modeling,DBLP:conf/hotpar/SaadMR12,DBLP:conf/systor/SaadPR18}.
In these cases when a predefined order is necessary, conflicts are handled by aborting (and re-executing) whichever transaction ran code with the latest chronological ordering. The key idea is that code blocks run as transactions and commit in the program's original chronological order. The techniques for supporting the aforementioned ordering are classified as: blocking~\cite{gonzalez2014effective, pldi09, von2007implicit} or freezing~\cite{zhang2010software}; a comparison between them is in~\cite{DBLP:conf/ppopp/SaadPR16}.

In the blocking approach, Mehrara \emph{et al.}~\cite{pldi09} proposed STMLite, a TM with a separate thread, Transaction Commit Manager (TCM), that detects conflicts among transactions waiting to commit.
TCM orchestrates the in-order commit process with the ability to have concurrent commits. Worker threads poll and stall to wait for the TCM's permission. Gonzalez \emph{et al.}~\cite{gonzalez2014effective} use a distributed approach for handling the commitment order. Each thread employs a bounded circular buffer to store its completed transactions. If all buffer slots are exhausted, the thread stalls until one of the pending executed transactions reaches the correct commit order.

Another existing solution lets threads \emph{freeze} completed transactions and proceed to execute the transactions with later chronological order, with the disadvantage of increasing the transaction lifetime (hence, a higher conflict probability). 
Zhang \emph{et al.}~\cite{zhang2010software} introduced this technique to support a predetermined total order of transactions. A \emph{next-to-commit} shared variable is used to preserve this order.
Overall, in both the blocking and freezing approach ordering transactions' commits negatively affects the overall resource utilization and may nullify any potential gain due to threads' parallelism. To overcome this restriction, OWB and OUL limit the stalling periods to only the latency of the commit.

The level of atomicity could be an orthogonal classification for the aforementioned techniques. The classical TM model mandates transactions to see only \emph{committed} values. However, concurrent transactions can construct a \emph{dependency graph} of uncommitted values. Based on this graph, the transactions commit in the constructed order. Ramadan \textit{et al.}~\cite{ramadan2009committing} proposed a dependency aware transactions model (DASTM), in which every object keeps track of all transactional reads and writes, and transactions forward their uncommitted changes to other conflicting transactions. Based on these relations, the commit order is defined at run-time by verifying that the constructed conflict graph is acyclic. This check is very expensive, especially when executed during the transaction execution and leads to performance gain only in the presence of very high conflicts.
As opposed to DASTM, OWB and OUL, and OUL-Steal do not maintain the conflicting graph because data forwarding is optimized to enforce only the predefined commit order.

Deterministic execution of TM may be seen as a distant related topic. Recently, in~\cite{Ravichandran:2014:DHD:2628071.2628094} it has been proposed an STM implementation that improves performance in case of deterministic execution. Deterministic execution is meant for reducing the possible parallelism in the system, whereas our approaches aim at introducing parallelism when a specific commit order is enforced.

Since code parallelization is the main application of our STM implementations, Thread-level Speculation (TLS)~\cite{Prabhu:2003:UTS:781498.781500,Oancea:2009:LII:1583991.1584050} is an immediate related topic. Loop parallelization using TLS has been proposed in both hardware~\cite{Prabhu:2003:UTS:781498.781500} and software~\cite{Rauchwerger:1995:LTS:223428.207148,Bhowmik:2002:GCF:564870.564885}. TLS and TM have been merged through a unified model in~\cite{barreto2012unifying,ramaseshan2008toward, Raman:2010:SPU:1736020.1736030,Oancea:2009:LII:1583991.1584050} to get the best of the two techniques. Generally, TLS is a less flexible way to parallelize code than using STM. For example, with STM only some instructions can be instrumented while the others still execute without instrumentation; on the contrary, leveraging TLS means speculating over the entire loop body. 
%Transactional Memory (TM) is an abstraction (thus more flexible) that allows for finer-grained speculation (e.g., not all instructions in a loop must be included in the transaction). One example is the way we envision the usage of our algorithms, namely integrated in a system for parallelization where transactional instrumentation is the minimum required to preserve original application semantics. Our algorithms will then execute these transactions in the pre-defined order. 
To overcome some of the well-known TLS limitations, the work in~\cite{Oancea:2009:LII:1583991.1584050} proposes a software TLS implementation where write operations directly update the non-speculative memory and read-races are tracked using metadata. Some of these intuitions have been ported to STM by OWB, OUL, and OUL-Steal.

\section{Execution and Memory Model}
\label{sec:tx-model}

Our model assumes a set of transactions $\{ T_1, T_2, \ldots, T_N \}$. Transactions access shared objects using read and write operations, with their usual meaning~\cite{herlihy93transactional}. We denote the sets of shared objects accessed by transaction $T_i$ for read and write as \texttt{read-set($T_i$)} and \texttt{write-set($T_i$)}, respectively.

\begin{figure}[h]
\centering
\includegraphics[scale=0.35]{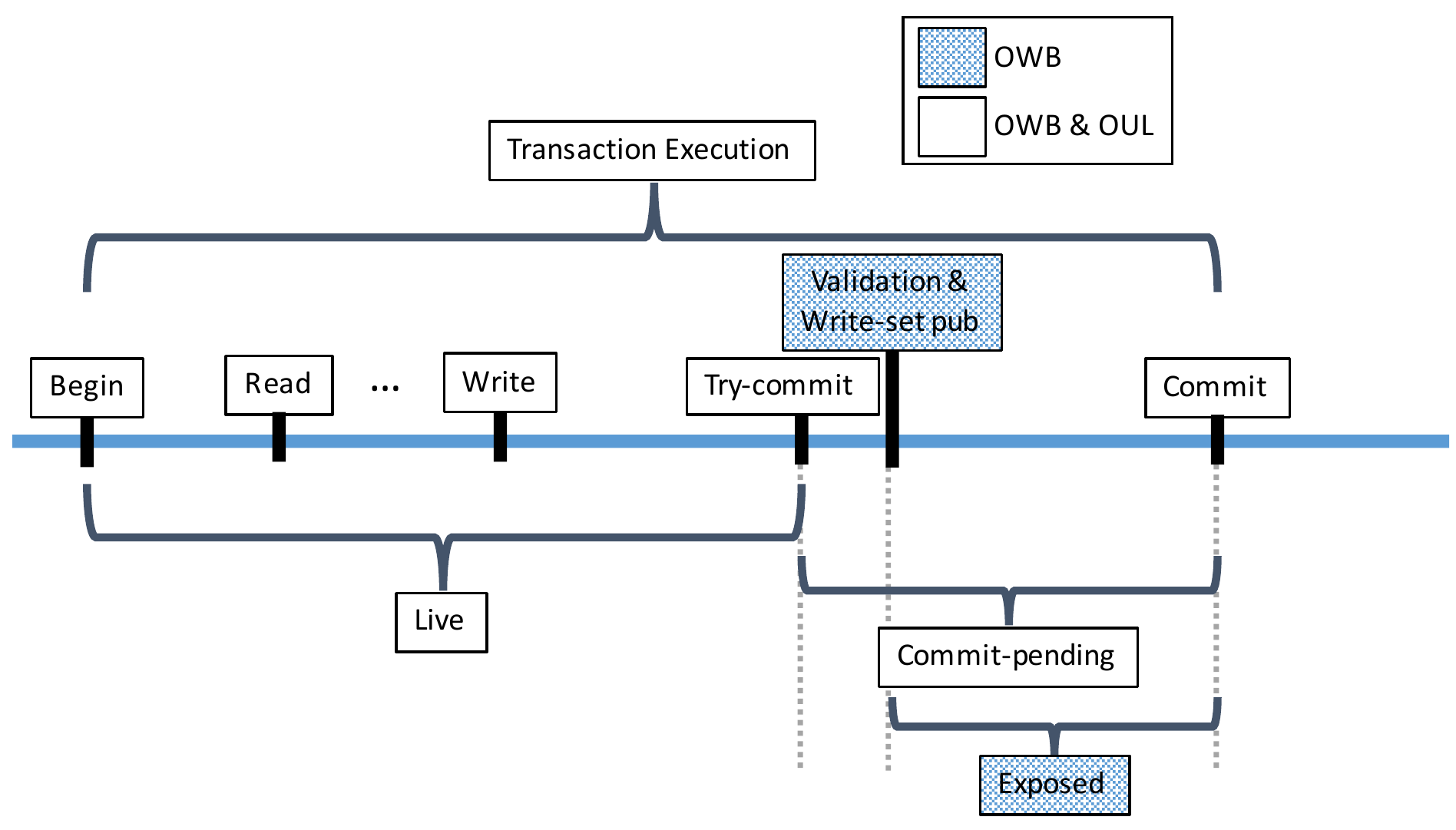}
\caption{Transaction execution states in OWB and OUL.}
\label{fig:tx-states}
\end{figure}

A transaction execution is defined as a sequence of operations, where each operation is represented by a pair of \textit{invoke} and \textit{return} events. Besides the read and write operations, whose semantics is the usual one, it also includes a commit operation that starts by invoking the \textit{try-commit} event, whose return value is either \textit{commit} or \textit{abort}. Note that a transaction can also be aborted before invoking the try-commit event. A transaction that begins its execution and did not invoke the try-commit event yet is called \textit{live}. A transaction that invoked the try-commit event but did not commit or abort yet is called \textit{commit-pending}. 
When a transaction is categorized as committed, it means that all its write operations have been executed permanently on the shared state; and when it is categorized as aborted, its operations have no permanent effect. In both the cases, all metadata is cleaned before proceeding or re-executing. Figure~\ref{fig:tx-states} summarizes the transaction states.

A shared object has a value and a (versioned) lock associated with it. We say that a shared object is \textit{exposed} if it is locked by some live or commit-pending transaction. Intuitively, a shared object is exposed if some transaction can already read it, although the transaction that wrote to that object is still executing. A transaction is \textit{exposed} when it is commit-pending and has all its written objects exposed.

Two transactions are said to be \emph{conflicting} if both are concurrent and access an object $X$, and at least one of them writes on $X$. Note that two transactions are conflicting even if both write the same object without reading it. Including such a dependency is fundamental, as motivated in the next paragraph. A conflict is handled by aborting one of the transactions, or postponing the access responsible for the conflict (if possible), until the other transaction commits.

\textbf{Transaction Ordering.}
We focus on TM implementations providing a specific order of transaction commits, which is assumed to be known prior to the transaction execution.
We denote such an order as the transaction \emph{age}.
The age is assumed to be defined before activating any transaction (e.g., an ordering layer deployed on top of the TM implementation), and must match the transactions commit order. The age is \textit{unique}, meaning no two transactions can have the same age and once assigned to a transaction, the age does not change even if the transaction is aborted multiple times.

The age of a transaction should not be confused with the transaction timestamp taken from a global timestamp, as used by many existing TM implementations (e.g., TL2~\cite{Dice06transactionallocking}, LSA~\cite{riegel:lsa:disc:2006}). The global timestamp is advanced by the concurrency control every time a transaction commits. The age of a transaction is externally determined (e.g., by the application) and does not depend on the execution of concurrent transactions.

Let $\prec$ be the total order relation on transaction ages, and let those ages be denoted as subscripts (e.g., $T_i$).
If $T_i \prec T_j$ we say that $T_i$ has a lower age than $T_j$; otherwise higher. A concurrency control that enforces an order of commits ensures that when two operations $o_i$ and $o_j$, issued by transactions $T_i$ and $T_j$, respectively, are conflicting, then $o_i$ must happen before $o_j$ if and only if $T_i \prec T_j$. We deploy this idea into our execution model by introducing an Age-based Commit Order (ACO). ACO mandates a customization of the classical TM model. As an example, the transaction conflict detection should guarantee that when $T_i \prec T_j$, $T_i$ must not read a value written by $T_j$ (intuitively, $T_i$ should commit before $T_j$).

We define a transaction $T_j$ as \emph{reachable} if all $T_j$'s lower age transactions are committed, which means that $T_j$ has been reached by a serial execution where all transactions \{$T_1$,\dots, $T_{j-1}$\} committed in the order $1,\dots,j-1$.
In practice, ACO constrains the serialization orders.

In our model, when a transaction aborts, it is restarted by the TM library with the same age.

\section{General Design}
\label{seq:ooo}
In this section, we present our co-operative model for supporting ACO. The core idea is to relax the common practice of letting transactions access values written by only committed or commit-pending transactions that will surely commit. In our proposed solutions, we weaken this assumption while still preserving the consistency according to ACO.
Depending on the desired correctness and performance level, we permit a transaction to expose its changes either:
\begin{compactitem}[-]
\item after it invokes the try-commit event and performs a validation to verify execution's consistency, but still allowing it to abort later due to ACO violation (in OWB); or
\item right after the write operation takes place during the execution and aborting any dependent transaction as soon as a further modification on the same early exposed value happens (in OUL).
\end{compactitem}

The above idea allows transactions with higher age to use such visible changes. Although it speeds up the flow of values from lower to higher age transactions, it also creates a possible dependency chain with other live and commit-pending transactions that accessed those values. Therefore when an abort occurs, the abort event should be immediately triggered to all the dependent transactions (cascading abort).

We can classify existing transactional models according to the way they handle concurrent read and write operations as: conservative model, conflict serialization model (e.g., DASTM~\cite{ramadan2009committing}), and cooperative ordered model (our model). The former prohibits the co-existence of read and write operations on the same object issued by concurrent transactions (this model is deployed by most concurrency controls). The conflict serialization model permits all combinations and selects the commit order based on the transaction dependency. Our cooperative ordered model restricts the memory snapshot seen by transactions to only the values exposed by transactions with lower age.

Interestingly, under the conflict serialization model transactions are aborted only when a mutual dependency (i.e., a cycle in the transaction dependency graph) exists; in our model, the graph is always acyclic. Avoiding to identify cycles in the transaction dependency chain increases the chance to achieve high performance.

\subsection{Cooperative Ordered Transactional Execution}
\label{subsec:tx-ex-order}

To construct our cooperative model, we start by highlighting the following two events of a transaction execution:
\begin{compactitem}[-]
\item a transaction is \textit{exposed} when all its written objects are exposed, and it is in the commit-pending state by having all its read operations consistent according to the ACO, therefore no conflict with lower age transactions occurred;

\item a transaction becomes \textit{reachable}, when all the lower age transactions have been committed.

\end{compactitem}

Supporting this new model requires that: \textit{i)} aborted transactions should be able to abort other transactions that accessed their exposed updates; and \textit{ii)} lower age transactions should enforce the abort of exposed higher age transactions.
Accomplishing the above goals requires maintaining some transactional metadata (e.g., read and write sets, including acquired locks) even after a transaction is exposed. Those metadata help in identifying conflicts (or aborting) exposed transactions, and they should be kept accessible until a transaction becomes reachable. Additionally, we need to support the cascading abort of multiple live or exposed transactions that share elements in their read-sets and write-sets.

Exposing written objects before being sure that a transaction eventually commits may violate the ACO if all lower age transactions are not committed yet. Similarly, ACO might be violated when the transaction conflicts with a lower age transaction that accesses a common object that is exposed by the first. For this reason, we postpone releasing the transaction metadata until the transaction becomes reachable, thus providing a safe point to decide whether a commit or abort should be triggered. The main difference between an exposed transaction and a reachable transaction is that: the former, although it has already published its modifications, it can still be aborted (and trigger the cascading abort of other transactions), whereas the latter cannot be aborted anymore. Therefore, it is safe to release all its metadata without violating the ACO.

\section{The Ordered Write Back (OWB)}
\label{sec:owb}

\begin{algorithm*}[h]
\caption{OWB - Pseudocode}
\label{owb-algo-rw}
\scriptsize
\begin{algorithmic}[1]
\begin{multicols}{2}
\Procedure{Read(SharedObject so, Transaction tx)}{}
\State readVersion = so.lock.version \label{restart_read}
\State currentWriter = so.lock.writer
\If {tx.writeSet.contains(so)}
	\State \Return tx.writeSet.get(so).value		\Comment{Read written value}
\ElsIf {currentWriter $\neq$ NULL}				\Comment{Check speculative write}
	\If {currentWriter.age $\gt$ tx.age}
        \State ABORT(currentWriter)	\Comment{$W_2 \to R_1$; Read after Speculative Write} 
        \Goto{restart_read}
    \Else	\Comment{$W_1 \to R_2$; Add Tx to its dependencies}
		\State currentWriter.dependencies.add(tx)
		\If {currentWriter.status $\ne$ ACTIVE}	\Comment{Double check writer}
			\State ABORT(tx)	\Comment{Writer got aborted while registeration}
		\EndIf
	\EndIf
\EndIf
\State Validate\_Reads(tx)
\State tx.readSet.add(so, readVersion)
\State \Return so.value
\EndProcedure

\

\Procedure{Write(SharedObject so, Object value, Transaction tx)}{}
\State tx.writeSet.add(so, newValue)		\Comment{Save new value}
\EndProcedure

\

\Procedure{Abort(Transaction tx)}{}
\IIf {tx.status = ABORTED} \Return false; \EndIIf \Comment{Already got aborted} \label{abort_checkstate}
\IIf {tx.status = INACTIVE} \Return false; \EndIIf \Comment{Already compeleted}
\While { ! CAS(tx.status, ACTIVE, TRANSIENT) }	\Comment {Try Abort}
	\Spin { tx.status $\neq$ TRANSIENT }	\Comment {Spin Wait}
	\Goto {abort_checkstate}
\EndWhile
\For{each Transaction $dependency$ in tx.dependencies}
	\State ABORT(dependency)		\label{owb-cascabort} \Comment {Abort dependent transactions}
\EndFor
\For{each Entry $entry$ in tx.writeSet}
	\State SharedObject so = entry.so
	\If {so.lock.writer $=$ tx}		\Comment {Aquired lock}
%		\State so.lock.version - - 
		\State so.value = entry.newValue	\Comment {Revert value}
		\State so.lock.writer = NULL		\Comment {Release the lock}
	\EndIf
\EndFor
\State tx.status = ABORTED
\State \Return true
\EndProcedure

\

\Procedure{Validate\_Reads(Transaction tx)}{}
\For{each Entry $entry$ in tx.readSet}	\Comment {Validate Read Set}
	\State SharedObject so = entry.so
	\If {so.lock.version $\neq$ entry.readVersion $\wedge$ so.lock.writer $=$ NULL}
		\State \Return ABORT(tx)	\Comment {Read a wrong version}
	\EndIf
\EndFor
\State \Return VALID
\EndProcedure

\

\Procedure{Validate\_Locked\_Reads(Transaction tx)}{}
\For{each Entry $entry$ in tx.readSet}	 \Comment {Validate Write Set}	
	\State SharedObject so = entry.so
	\If {so.lock.writer $=$ tx $\wedge$ so.lock.version $\neq$ 1 + entry.readVersion} 
		\State \Return ABORT(tx)	\Comment {Concurrent Expose/Commit Occurs} 
	\EndIf
\EndFor
\State \Return VALID
\EndProcedure

\

\Procedure{TryCommit(Transaction tx)}{}
\IIf {tx.status = ABORTED} \Return false; \EndIIf \Comment{Already got aborted}
\While { ! CAS(tx.status, ACTIVE, TRANSIENT) }	\Comment {Try Commit}
	\Spin { tx.status $\neq$ TRANSIENT }	\Comment {Spin Wait}
	\State \Return false
\EndWhile
\IIf {VALIDATE\_READS(tx) $\neq$ VALID}	\Return false;	\EndIIf \label{validate-read}
\For{each Entry $entry$ in tx.writeSet}	\Comment {Lock Write Set}
	\State SharedObject so = entry.so
	\State currentWriter = so.lock.writer \label{marker_vread}
	\If {currentWriter $\neq$ NULL}
		\If {tx.age $\lt$ currentWriter.age}
			\State ABORT(currentWriter)	\Comment{$W_2 \to W_1$; Write after Specu. Write}
		\Else
			\State ABORT(tx)		\Comment{$W_1 \to W_2$; Write after Write}
			\State \Return false
		\EndIf
		\If { ! CAS(so.lock.writer, NULL, tx) } \Comment {Acquire Lock}
			\Goto{marker_vread}
		\EndIf
	\EndIf
\EndFor
\For{each Entry $entry$ in tx.writeSet}
	\State SharedObject so = entry.so
	\State so.lock.version + + 
	\State temp = so.value				\Comment {Save old value} \label{swap1}
	\State so.value = entry.newValue	\Comment {Expose written value}
	\State entry.newValue = temp							\label{swap2}
\EndFor
\IIf {Validate\_Locked\_Reads(tx) $\neq$ VALID}  \Return false; \EndIIf	\label{validate-locked}
\State tx.status = ACTIVE	\Comment {Transaction Exposed}
\State \Return true
\EndProcedure

\

\Procedure{Commit(Transaction tx)}{}
\IIf {tx.status = ABORTED} \Return false; \EndIIf \Comment{Already got aborted}
\While { ! CAS(tx.status, ACTIVE, TRANSIENT) }	\Comment {Try Complete}
	\Spin { tx.status $\neq$ TRANSIENT }	\Comment {Spin Wait}
	\State \Return false
\EndWhile
\IIf {VALIDATE\_READS(tx) $\neq$ VALID} \Return false; \EndIIf
\For{each Entry $entry$ in tx.writeSet}
	\State SharedObject so = entry.so
	\State so.lock.writer = NULL	\Comment {Unlock}
\EndFor
\State tx.status = INACTIVE	\label{owb-commit}	\Comment {Transaction Committed}
\State \Return true
\EndProcedure

\end{multicols}
\end{algorithmic}
\end{algorithm*}

The Ordered Write Back Algorithm (\textit{OWB}) employs a write-buffer approach; a transaction writes its modifications into a local buffer. While entering the try-commit phase, the transaction acquires a versioned-lock for each object in its write-set and writes its changes to the shared memory, and becomes exposed. To avoid concurrent writers, the locks are not released until the transaction becomes committed or is aborted.
However, to allow an early propagation of the modifications, higher age transactions can read these locked objects. In case an abort is triggered, the exposed transaction is responsible to abort any dependent transaction that has read the exposed values.
We use versioning to detect conflicts between concurrent transactions. The transaction performs a validation before exposing its values, and before releasing its locks to approach the final commit.

In practice, for OWB a transaction is exposed if: it is executed until the end without any conflict with other concurrent transactions; it acquired the locks on its modified objects successfully; it exposed their new values to the shared memory; and it is waiting to be reachable. A transaction can commit only if it is reachable and passes the validation of its read operations. The transaction also releases its acquired locks at this stage.
As stated earlier, an exposed transaction can still be aborted.

A transaction keeps these metadata: \textit{1)} \emph{read-set}, which stores read objects and their read version; \textit{2)} \emph{write-set}, which stores the modified objects and their new values; and \textit{3)} \emph{dependencies list}: a list of transactions who read the changes done by this transaction after it becomes exposed.
Shared objects are associated with a versioned lock. The lock stores the version number and a reference to the writer transaction (if it exists) that currently owns it. The version is incremented when a new value for the object is exposed. The pseudocode of OWB is in Algorithm~\ref{sec:owb}.

The \texttt{Write} operation simply adds the object and its new value to the write-set.
The \texttt{Read} operation first checks if the object has been earlier modified by the transaction itself. If so, the new value from the write-set is returned;
otherwise the object, along with its version, is fetched from the shared memory. If the object is currently exposed, then the writer is aborted only if its age is higher ($W_2 \to R_1$), and the read operation is retried.
If the transaction that holds the lock has a lower age than the reading transaction, we let the latter read the written value ($W_1 \to R_2$) and add itself to the writer's dependencies list. That way, if the writer will be aborted in the future, it can cascade its abort to the affected transactions who read its modified objects.
It is worth noting that, to avoid inconsistencies while reading from an exposed writer, we let the reader double check the writer state (if it is aborted) after it registers itself in the writer dependencies list; also the dependencies list must provides a thread-safe insertion.
Before a read operation returns, the read-set is validated by invoking \texttt{Validate\_Reads} (see below). Not doing that would make the read-set of OWB transactions not consistent during execution.

To enter the exposed state, a validation of the read-set is executed to make sure the transaction reads a consistent view of the memory before exposing the locally buffered written objects. \texttt{Validate\_Reads} compares the current versions of the read objects with the value of the corresponding versions stored in the read-set.
If the current version is different, then this means that the object was modified after the read, i.e., a Write after Read (WAR) conflict.

Upon passing a read-set validation, the \texttt{exposed} procedure acquires the locks and then writes the write-set to the memory. If the locks are already acquired by another concurrent exposed writer ($W_1 \to W_2$ or $W_2 \to W_1$), we handle that by favoring the lower age transaction, and aborting the other.
Since exposed transactions can still be aborted by other transactions, we need to store the old value of modified objects. This is done by swapping the write-set stored values with the old objects' values at commit.

Finally, at commit time we call \texttt{Validate\_Reads} again to prevent the WAR anomaly.
However, since write-set elements are already locked, we can leverage that to reduce the validation overhead. Consider $T_i$ is executing the commit operation. Let $X \in$ \texttt{read-set}($T_i$) $\cap$ \texttt{write-set}($T_i$). As $T_i$ is still acquiring the locks over its write-set (including $X$), $T_i$ is sure that the value of $X$ is unchanged since its lock acquisition, thereby it could be excluded from the commit-time read-set validation. To do so, it requires checking that read-set objects have not been changed while acquiring locks.

Keeping the commit execution time short is fundamental; therefore, the optimization just described shrinks the commit execution at the price of adding an extra check in the \texttt{Try-Commit} procedure. However, having an object read and written in a transaction is a common pattern, which makes this optimization fruitful.

Finally, when a transaction becomes reachable and the re-validation succeeds, the \texttt{commit} operation releases its acquired locks and reclaims metadata.

As correctness guarantee, OWB guarantees TMS1~\cite{DohertyGLM2013}. The intuition is that: if for a history generated by OWB, every exposed transaction is committed, then the history is opaque~\cite{gue08}. 
First of all, transactions can commit only in the ACO order, serializing all the committed transactions, making OWB strict serializable.
Moreover, OWB allows transactions to read only from commit-pending (exposed) and committed transactions, and any time a transaction enters the exposed state, it aborts all concurrent transactions that has read a value that violates the ACO. However, exposed transactions can abort after some live transaction already read those values. This is not allowed by Opacity, but TMS1 allows that as long as the live transactions do not perform any operation after the exposed transaction is aborted. OWB implements that through an atomic cascading abort. We give more details about correctness in Section~\ref{sec:correctness}.

\section{The Ordered Undo Log (OUL)}
\label{sec:oul}

The Ordered Undo Log (\textit{OUL}) Algorithm is an undo-log algorithm that preserves the ACO. Here, transactional updates affect the shared memory at encounter time, while the old value is kept in a local undo-log.
Such a scheme implies that the transactions' order is guaranteed while operations are invoked, and not at commit time as in OWB.
In order to deploy the above idea, each object is associated with a read-write lock. The transaction acquires a read or write lock according to its need, as explained later.
Also, each lock stores the reference to the (single) writer transaction, which can be either the current transaction holding the lock or the one that committed that version,
and a list of concurrent readers, namely those transactions that accessed the version for reading it, and they are still live or commit-pending.
Note that the size of the readers list impacts the efficiency of the protocol, thus it should be bounded.

As in OWB, every transaction in OUL maintains a write-set, but here the write-set stores the old values of the written objects (undo-log). Regarding the transaction read-set, it is implicitly represented by the object lock's readers list.

The pseudocode of OUL's core operations is included in Algorithm~\ref{oul-algo} and Algorithm~\ref{oul-algo-2}.
In the \texttt{Read} procedure, we allow Read after Write (RAW) conflicts only if the writer transaction has a lower age ($W_1 \to R_2$); otherwise the speculative writer is aborted ($W_2 \to R_1$).
The \texttt{Write} procedure enforces that only a single transaction can hold the write lock on the object at a time.
A Write-Write conflict is solved by aborting the transaction with the highest age. As readers are visible, the writer transaction can check if there is any (wrong) speculative reader, and abort it accordingly ($R_2 \to W_1$).

One of the major benefit of a write through protocol is that the \texttt{Try-Commit} procedure is simple because the values are already in the shared memory. However, in OUL exposing a transaction only means that it did not conflict with other transactions so far -- but it could be still aborted to preserve the ACO.
In the \texttt{Commit} procedure, the transaction is marked as \texttt{Inactive} and locks are released. As we said before, since a lock is maintained with a back-reference to the transaction that holds it, setting the transaction status is sufficient to release all the locks held by that transaction with a single step. On the other hand, in the \texttt{Abort} procedure the transaction restores old values from the undo-log (Rollback), and release all the locks (switches to \texttt{Inactive}).

\begin{algorithm*}[h]
\caption{OUL - Pseudocode 1}
\label{oul-algo}
\scriptsize
\begin{algorithmic}[1]
\Procedure{Read(SharedObject so, Transaction tx)}{}
\IIf {tx.status = TRANSIENT}	\Return ABORT(tx)	\EndIIf \label{retry_read} 
\State Transaction currentWriter = so.lock.writer;
\IIf {currentWriter = BUSY }	\IGoto{retry_read}	\EndIIf
\If {currentWriter $\neq$ NULL $\wedge$ currentWriter.status $\neq$ INACTIVE $\wedge$ currentWriter.age $\gt$ this.age}
    \State ABORT(currentWriter)	\Comment{$W_2 \to R_1$; Read after specu. Write}
    \Goto{retry_read}
\EndIf
\State registered = false
\Repeat
	\For{i=0 to $MAX\_READERS$}
		\State Transaction readerSlot = so.lock.reader[i]
		\If {readerSlot $\neq$ ACTIVE $\wedge$ readerSlot $\neq$ PENDING $\wedge$ CAS(so.lock.reader[i], readerSlot, tx)}
			\State registered = true					\Comment{Found empty reader slot}
		\EndIf
	\EndFor
\Until {registered}
\If {currentWriter $\neq$ so.lock.writer}	\Comment{Writer was changed}
    \Goto{retry_read}
\EndIf
\State \Return so.value
\EndProcedure

\Procedure{Write(SharedObject so, Object value, Transaction tx)}{}
\IIf {tx.status = TRANSIENT} ABORT(tx); \EndIIf	\label {retry_write}
\State Transaction currentWriter = so.lock.writer;
\IIf {currentWriter = BUSY }	\IGoto{retry_write}	\EndIIf
\If {currentWriter $\neq$ tx }	\Comment{ Already in write-set}
	\If {currentWriter $\neq$ NULL $\wedge$ currentWriter.status $\neq$ INACTIVE}
		\If {currentWriter.age $\gt$ this.age }
	    	\State ABORT(currentWriter)	\Comment{$W_2 \to W_1$; Write after specu. Write}
    		\Goto{retry_write}
	    \Else
	    	\State ABORT(tx)	\Comment{$W_1 \to W_2$; Write after Write}
	    \EndIf
	\EndIf
	\If { ! CAS(so.lock.writer, currentWriter, BUSY) }	
	    	\Goto{retry_write} \Comment{Failed to aquire the lock}
	\EndIf
	\State tx.writeSet.add(so, so.value)		\Comment{Save old value}
\EndIf	
\For{i=0 to $MAX\_READERS$}
	\State Transaction readerSlot = so.lock.reader[i]
	\If {readerSlot $\neq$ INACTIVE $\wedge$ readerSlot.age $\gt$ tx.age)}  \label{abort_readers}
		\State ABORT(readerSlot)	\Comment{$R_2 \to W_1$; Abort specu. reader}
	\EndIf
\EndFor
\State so.value = newValue					\Comment{Write new value}
\State so.lock.writer = tx					\Comment{Save me as the new writer}
\EndProcedure
\end{algorithmic}
\end{algorithm*}

\begin{algorithm}[h]
\caption{OUL - Pseudocode 2}
\label{oul-algo-2}
\scriptsize
\begin{algorithmic}[1]
\setcounter{ALG@line}{49}
\Procedure{TryCommit(Transaction tx)}{}
\IIf { !CAS(tx.status, ACTIVE, PENDING)} ABORT(tx) \EndIIf
\EndProcedure
\Procedure{Commit(Transaction tx)}{}
\IIf { CAS(tx.status, PENDING, INACTIVE)} \Return true  \label{oul-commit}	\EndIIf
\Spin {tx.status $\neq$ TRANSIENT}
\Comment{Wait until be aborted}
\EndProcedure

\Procedure{Abort(Transaction tx)}{}
\IIf {tx.status = INACTIVE} \Return true;	\EndIIf	

\Comment {Check if already aborted}
\If {CAS(tx.status, PENDING, TRANSIENT)}
	\Comment {Rollback}
	\For{each Entry $entry$ in tx.writeSet}
		\State SharedObject so = entry.so
		\State Object value = entry.value
		\State so.value = value		\Comment {Restore old value}
		
		\For{i=0 to $MAX\_READERS$}
			\State Transaction readerSlot = so.lock.reader[i]
			\If {readerSlot $\neq$ INACTIVE $\wedge$ readerSlot.age $\gt$ tx.age)}
				\State ABORT(readerSlot) \Comment{Abort specu. reader}
			\EndIf
		\EndFor
	\EndFor
	\State tx.status = INACTIVE
\Else 		\Comment {Set aborted}
	\State \Return {CAS(tx.status, ACTIVE, TRANSIENT)}
\EndIf
\EndProcedure
\end{algorithmic}
\end{algorithm}

\begin{algorithm*}[h]
\caption{OUL-Steal - Pseudocode}
\label{oul-steal-algo}
\scriptsize
\begin{algorithmic}[1]
\begin{multicols}{2}

\Procedure{Read(SharedObject so, Transaction tx)}{}
\IIf {tx.status = TRANSIENT}	\Return ABORT(tx)	\EndIIf \label{oul-stel-retry_read} 
\State Transaction currentWriter = so.lock.writer;
\IIf {currentWriter = BUSY }	\IGoto{oul-stel-retry_read}	\EndIIf
\If {currentWriter $\neq$ NULL $\wedge$ currentWriter.status $\neq$ INACTIVE $\wedge$ currentWriter.age $\gt$ this.age}
    \State ABORT(currentWriter)	\Comment{$W_2 \to R_1$; Read after Speculative Write}
    \Goto{oul-stel-retry_read}
\EndIf
\State registered = false
\Repeat
	\For{i=0 to $MAX\_READERS$}
		\State Transaction readerSlot = so.lock.reader[i]
		\If {readerSlot $\neq$ ACTIVE $\wedge$ readerSlot $\neq$ PENDING $\wedge$ CAS(so.lock.reader[i], readerSlot, tx)}
			\State registered = true					\Comment{Found empty reader slot}
		\EndIf
	\EndFor
\Until {registered}
\If {currentWriter $\neq$ so.lock.writer}	\Comment{Writer got changed meanwhile}
    \Goto{oul-stel-retry_read}
\EndIf
\State \Return so.value
\EndProcedure

\

\Procedure{Write(SharedObject so, Object value, Transaction tx)}{}
\IIf {tx.status = TRANSIENT} ABORT(tx); \EndIIf	\label{steal_retry_write}
\State Transaction currentWriter = so.lock.writer
\IIf {currentWriter = BUSY } \IGoto{steal_retry_write} \EndIIf
\If {currentWriter $\neq$ tx }	\Comment{ Already in write-set}
	\State steal = false
	\If {currentWriter $\neq$ NULL $\wedge$ currentWriter.status $\neq$ INACTIVE}
		\If {currentWriter.age $\gt$ this.age }
	    	\State ABORT(currentWriter)		\Comment{$W_2 \to W_1$; Write after Specu. Write}
    		\Goto{steal_retry_write}
	    \Else
	    	\State steal = true	\Comment{$W_1 \to W_2$; Lock Steal, Write after Write} \label{steal1}
	    \EndIf
	\EndIf
	\If { ! CAS(so.lock.writer, currentWriter, BUSY) }	\Comment{Aquire the lock}
    		\Goto{steal_retry_write}
	\EndIf
	\State tx.writeSet.add(so, so.value, steal ? currentWriter : NULL)		\Comment{Save old value, and old writer when stealing the lock} \label{steal2}
\EndIf
\For{i=0 to $MAX\_READERS$}
	\State Transaction readerSlot = so.lock.reader[i]
	\If {readerSlot $\neq$ INACTIVE $\wedge$ readerSlot.age $\gt$ tx.age)}
		\State ABORT(readerSlot)	\Comment{$R_2 \to W_1$; Abort speculative readers}
	\EndIf
\EndFor
\State so.value = newValue					\Comment{Write new value}
\State so.lock.writer = tx					\Comment{Save me as the new writer}
\EndProcedure

\

\Procedure{TryCommit(Transaction tx)}{}
\IIf { !CAS(tx.status, ACTIVE, PENDING)} ABORT(tx) \EndIIf
\EndProcedure

\

\Procedure{Commit(Transaction tx)}{}
\IIf { CAS(tx.status, PENDING, INACTIVE)} \Return true  \EndIIf
\Spin {tx.status $\neq$ TRANSIENT}	\Comment{Wait until be aborted}
\EndProcedure

\

\Procedure{Rollback(Transaction tx)}{}
	\State tx.aborted = true
	\For{each Entry $entry$ in tx.writeSet}
		\State SharedObject so = entry.so
		\If {CAS(so.lock.writer, tx, BUSY)}
			\State Object value = entry.value
			\State so.value = value		\Comment {Restore old value}
			\State so.lock.writer = entry.originalOwner		\Comment {Release lock, or return the original owner}
			\If { entry.originalOwner != NULL $\wedge$ entry.originalOwner.aborted}
				\State ROLLBACK(entry.originalOwner)	\label{steal-rollback}
			\EndIf
		\EndIf
		\For{i=0 to $MAX\_READERS$}
			\State Transaction readerSlot = so.lock.reader[i]
			\If {readerSlot $\neq$ INACTIVE $\wedge$ readerSlot.age $\gt$ tx.age)}
				\State ABORT(readerSlot)	\Comment{Abort speculative readers}
			\EndIf
		\EndFor
	\EndFor
\EndProcedure

\

\Procedure{Abort(Transaction tx)}{}
\IIf {tx.status = INACTIVE} \Return true; \EndIIf	\Comment {Already Aborted}
\If {CAS(tx.status, PENDING, TRANSIENT)}
	\State Rollback(tx)	\Comment {Rollback}	
	\State tx.status = INACTIVE
\Else 
	\If {CAS(tx.status, ACTIVE, TRANSIENT)}	\Comment {Set aborted}
		\State \Return true
	\EndIf
\EndIf
\State \Return false \Comment {Failed to abort} 
\EndProcedure
\end{multicols}
\end{algorithmic}
\end{algorithm*}

\subsection{The OUL-Steal Algorithm}
In this section, we introduce OUL-Steal, a variant of the OUL algorithm where we relax the aforementioned multiple-writers restriction and allow write-writer conflicts while guaranteeing ACO.
In both OWB and OUL, conflicting transactions co-operate to commit as they are allowed to proceed without aborts even in the presence of some read-write conflict, as long as ACO is still preserved.
However, a writer transaction holds the locks until reaching the \emph{commit} state, which sometimes limits the overall concurrency.

Let $T_i$ and $T_j$ be two conflicting writers on object $X$, and $T_i \prec T_j$.
In OUL, if $T_i$ finds $X$ locked by $T_j$ ($W_1 \to W_2$), $T_i$ should abort $T_j$.
However, ACO could still be preserved if $T_j$ overwrites the value written by $T_i$, as long as there is no other transaction $T_k$, with $T_i \prec T_k \prec T_j$, that will read $X$ in the future.

OUL-Steal allows a transaction with higher age to overwrite the value written by a concurrent transaction with lower age ($W_1 \to W_2$), and \emph{steal} its lock. The higher age transaction stores the stolen lock in a local list so that the lock can be returned back to the original writer (the lower age transaction) in case of abort.
That way, if a mid-age reader $T_k$ needs the value of a lower age transaction, then it can abort the higher age transaction(s) which stole the lock(s);
otherwise (i.e., without $T_k$), the value written by the higher age transaction will be used by higher age readers. 
This operation could be repeated until the reader reaches the correct writer transaction.

In \texttt{Write}, the lock is passed to the higher age writer and is saved in its write-set. As a consequence, the written address exists in the undo-log of both the writers (the original and the one which stole the lock). During the \texttt{Abort}, the transaction uses \texttt{Rollback} to revert its changes using its undo-log. An undo-log entry can be:
\begin{compactitem}[-]
\item \textit{stolen by} another writer: which means the transaction does not have the ownership record at the abort time. In this case, the transaction does not do any action, although, it keeps the undo-log entry, which contains the address value before the current transaction modifications.
\item \textit{exclusively modified} by the current transaction, reverting the old value from the undo-log, and aborting the speculative readers.
\item \textit{stolen from} another writer: in addition to the steps done in the \emph{exclusively modified} case, the lock ownership is passed back to the old writer, and the current transaction checks the state of the old writer. If it was \emph{not aborted}, then no further action is needed. Otherwise, the transaction calls the \texttt{Rollback} of the old owner. At this stage, the old writer will treat the entry as the cases of \textit{exclusively modified} or \textit{stolen from}, accordingly. 
\end{compactitem}
The complete pseudocode of the OUL-Steal algorithm is in Algorithm~\ref{oul-steal-algo}.

As correctness guarantee, OUL guarantees Strict Serializability~\cite{Papadimitriou79}. 
Unlike OWB, OUL allows reading from live transactions, which is not allowed by TMS1 (and hence opacity). 
However, similar to OWB, OUL restricts transactions to commit only in the ACO order, making OUL strict serializable.
More details about correctness are in Section~\ref{sec:correctness}.

\section{Correctness}
\label{sec:correctness} 
Here we discuss the correctness of the given algorithms. We do not include the case where a transaction triggers an error (e.g., division by zero) because it speculatively processes a computation that in a non-parallel execution would not happen. Such an execution might be a for-loop iteration executed speculatively and preceded by an iteration that breaks the for-loop. We assume a sandboxing mechanism to handle such exceptions.
%Another important observation related to the correctness of our algorithms is quiescence~\cite{mike-quicence,guy-of-the-call-and-last-ppopp19}. 

First, we show how our protocols preserve the ACO.
Suppose by contradiction that the ACO is violated. Let $T_i$ and $T_j$ be two transactions such that $T_i \prec T_j$. 
The interesting case is if $T_i$ successfully reads a value of an object $X$ written by $T_j$.
This implies that $R_i(X)$ happened after $T_j$ exposes X's value in OWB or  $write(T_j)$ in OUL.
In both OWL and OUL, $T_i$ acquires a shared lock on $X$ at the time of the read operation, either by visible reads (OUL, OUL-Steal) or checking if there is no writer (OWB). For a successful read, the shared lock must be acquired, thus the write lock should not be already granted. This implies that $T_j$ has released all its locks. As a transaction does not release its acquired locks until it \emph{commits}, $T_j$ must be necessarily committed. Therefore $R_i(X)$ must occur after $commit(T_j)$.
Since a transaction cannot perform any step after it commits, $R_i(X)$ $\to$ $commit(T_i)$.
This means $commit(T_j)$ $\to$ $commit(T_i)$, which cannot be the case since they must commit in order, according to their ages.

%\textbf{The correctness proofs of OUL's strict serializability and OWB's TMS1 are in the supplementary material.}

Now we prove that both OWB and OUL are serializable. In order to prove that, we define $DG(i, j)$ as a predicate defining a dependency between $T_i$ and $T_j$, when $T_j$ reads a value written by $T_i$, or $T_j$ overwrites a value written by $T_i$. Using this definition, we can construct a dependency directed graph $DG(T, D)$, where $T$ is the set of all committed transactions, and $D$ is the set of dependency relations. 
It is easy to see that $DG \subset SG$, where SG is the conflict serialization graph~\cite{bernstein1987concurrency}. A history is serializable if and only if its SG is acyclic. Note that serializability is not guaranteed if $DG$  is acyclic.

Assume by contradiction that an execution of our algorithms produce a cyclic $DG$, which implies having an edge $D(i, j)$ where $i > j$. By definition of dependency, this means that either $T_j$ reads a value written by $T_i$ (i.e., $W_i(X)$ $\to$ $R_j(X)$), or $T_j$ overwrites a $T_i$'s written value (i.e., $W_i(X)$ $\to$ $W_j(X)$). 
In all the proposed algorithms, exclusive locks must be acquired when we expose the written values (at commit in OWB or encounter time in OUL and OUL-Steal) and released only at commit, or passed to a higher age transaction (which is not the case here). We can rewrite the previous situations as $commit(T_i)$ $\to$ $R_j(X)$ or $commit(T_i)$ $\to$ $W_j(X)$. Since a transaction cannot perform any step after it commits,  $commit(T_i)$ $\to$ $commit(T_j)$, which cannot be the case as mentioned earlier; therefore, $DG$ is a acyclic.

Assume $e \in E = SG \setminus DG$, this edge represents the case where $R_i(X)$ $\to$ $W_j(X)$, which means $R_i(X)$ $\to$ $commit(T_j)$. In OWB, the procedure that validates read operations captures this by comparing the read version with the current version of the accessed object; while in OUL and OUL-Steal, the readers' visibility enables $T_j$ to detect the $R_i(X)$ and aborts it. So $E = \emptyset \implies SG = DG \implies SG$ is acyclic, making the algorithms serializable. 

The serialization point of both OUL and OWB is inside the commit procedure: for OUL is when the transaction's status is atomically set to \texttt{Inactive}; for OWB is when locks on written objects are released. As the serialization point is inside the transaction execution, all the algorithms preserve the real-time order, and are strict serializable.

In addition to being strict serializable, OWB is TMS1~\cite{DohertyGLM2013}, a stronger condition than strict serializability. 
Being TMS1, OWB ensures that response of every object operation, even by aborted and live transactions, is consistent with a serial execution.
Informally, for a history to be TMS1, it must be strict serializable, and for every successful response of an object operation by a transaction $T$, there must exist a serialization of a subset of the transactions, justifying the response.
This subset must contain $T$ (until the response) and all the committed transactions that completed before $T$ started.
In addition, the serialization can also contain some commit-pending transactions, and some committed and even aborted transactions, that are concurrent to $T$.
We have already shown earlier that OWB is strict serializable.
Since OWB allows a read operation to return a value written by an exposed transaction, which may get aborted later, it justifies including concurrent aborted transaction for the response. 
Recall that OWB allows reading values written by committed and exposed transactions only, but not from aborted transactions.
The intuition is that if a transaction reads from an exposed transaction, which gets aborted later, the reading transaction is also aborted without executing any further operations. 
This is done using cascading mechanism in OWB.

\section{Implementation and Evaluation}

In our implementation locks are implemented using 32 bits. The mapping between addresses and locks is made by leveraging the least significant bits, so a single lock might be responsible for multiple addresses.
The lock is divided into two parts: the most significant bits represent the reference to the writer, and the remaining bits represent either the header address of the readers list (for OUL), or the version number (for OWB). In OUL, we use a bounded list of readers to limit the number of concurrent readers, which is set to 40.

A thread plays multiple roles in our implementations: \emph{worker}, \emph{validator}, or \emph{cleaner}. A \emph{worker} executes transactions and performs the \texttt{try-commit}. A \emph{cleaner} takes care of reclaiming metadata. Once the transactional operations are all executed, any thread can take the lead of finalizing any transaction; however, there is a single thread at a time in the \emph{validator} role. This role is responsible for moving commit-pending transactions to the committed state and also re-executes invalid transactions. We adopt the \emph{flat combining}~\cite{hendler2010flat} technique to let threads take ownership of the \emph{validator} role. The pseudocode in Algorithm~\ref{thread-algo} shows the steps done by a thread to carry the execution of a transaction.

\begin{algorithm}[h]
\caption{Thread Execution}
\label{thread-algo}
\scriptsize
\begin{algorithmic}[1]

\For{each Transaction $tx$ in WorkQueue}
	\If {validator = IDLE $\wedge$ CAS(validator, IDLE, BUSY)} 
	
	\Comment {Try to be the validator}
		\State tx = ExposedList[last\_committed]	\label{retry-validation}
		\If { tx = NULL }			\Comment {Tx is not exposed yet}
			\Goto{stop-validation}			\Comment {Stop validation}
		\EndIf
		\If { tx.commit() = FAIL }	\Comment {Perform Tx commit}
			\State \ tx.start()
			\State \ tx.execute()		\Comment {Reexecute failed transaction}
			\State tx.tryCommit()		\Comment {Commit without validation}
			\State tx.commit()
		\EndIf
		\State last\_committed++
		\State CommittedQueue.enqueue(tx)
		\Goto{retry-validation}	\Comment {Validate next exposed Tx}
		\State validator = IDLE			\Comment {Release the validator role} \label{stop-validation}
	\EndIf
	\If {aborts$\gt$LIMIT $\vee$ tx.age - last\_committed $\gt$ MAX } \label{alarm}
		\While { tx.age - last\_completed $\gt$ MIN } \label{safe}
			\For{each Transaction $tx$ in CommittedQueue}
				\State tx.clean()	\Comment {Do housekeeping}
			\EndFor
		\EndWhile
	\EndIf	
	\State tx.start()	\label{retry-exec}
	\State tx.execute()	\Comment {Execute transaction}
	\If {tx.tryCommit() = FAIL}		\Comment {Try to expose transaction}
		\Goto{retry-exec}	\Comment {Retry}
	\EndIf
	\State ExposedList[tx.age] = tx	\Comment {Add to pending transactions}
\EndFor
\end{algorithmic}
\end{algorithm}

\begin{figure*}
\centering
	\begin{subfigure}[b]{\textwidth} \includegraphics[trim=1.5cm 4.3cm 0cm 0cm,clip=true,scale=0.6]{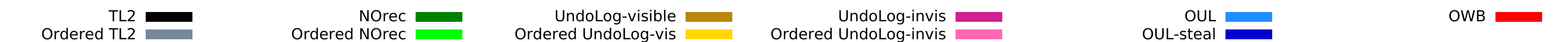}  \end{subfigure}
    \begin{subfigure}[b]{0.33\textwidth} \centering \includegraphics[scale=0.52]{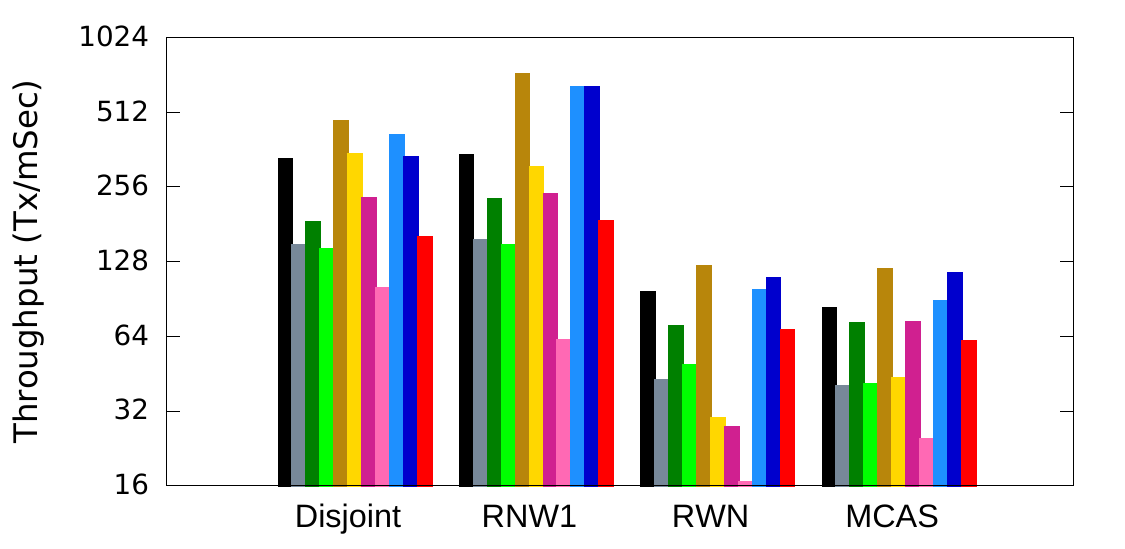} \vspace{-1.7em} \caption{Long Transaction} \label{fig:micro-all-long}  \end{subfigure}	
    \begin{subfigure}[b]{0.33\textwidth} \centering \includegraphics[scale=0.52]{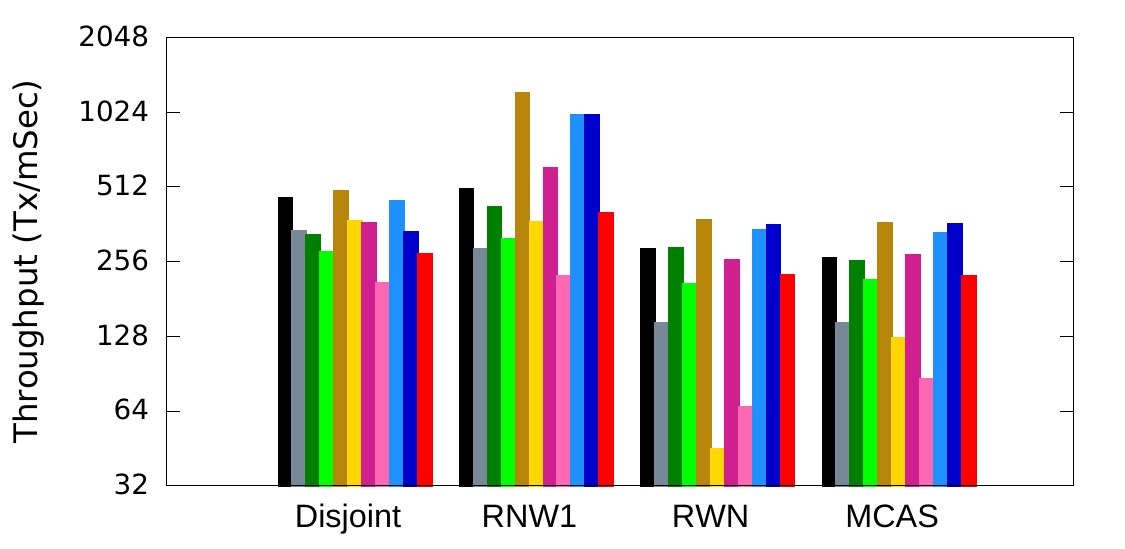} \vspace{-1.7em} \caption{Short Transaction} \label{fig:micro-all-short} \end{subfigure}
    \begin{subfigure}[b]{0.33\textwidth} \centering \includegraphics[scale=0.52]{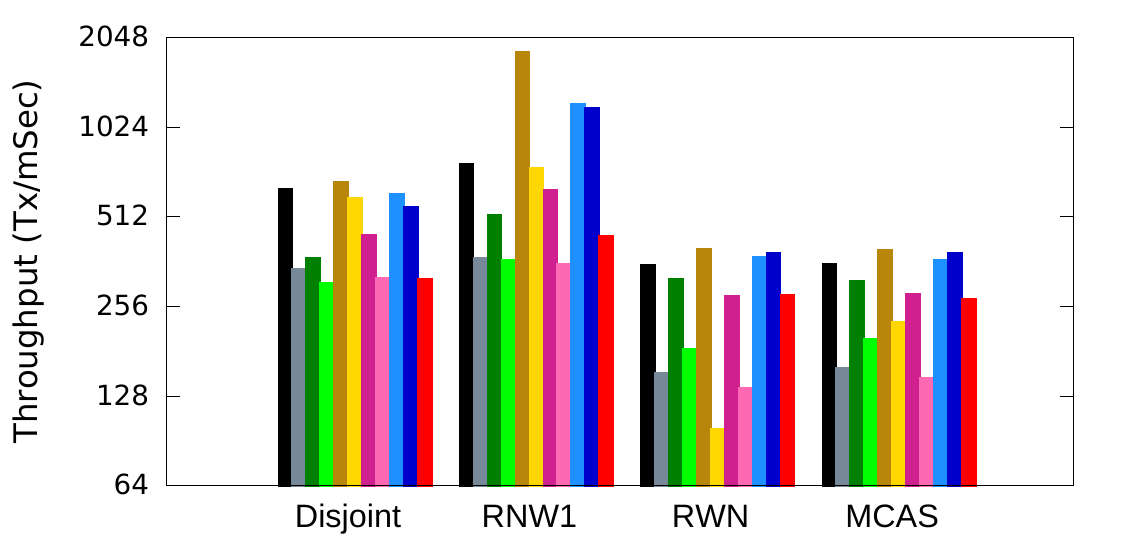} \vspace{-1.7em} \caption{Heavy Transaction} \label{fig:micro-all-heavy}  \end{subfigure}
    \vspace{-2em}
    \caption{Peak performance of all competitors (including unordered) using all micro benchmarks (Y-axis is log scale).} 
%    \vspace{-1.5em}
    \label{fig:micro-all}
\end{figure*}
We compare our algorithms with STMLite~\cite{pldi09}: a lightweight STM with ACO used to support code parallelization; the unordered and ordered version of three state-of-art TM algorithms: TL2~\cite{Dice06transactionallocking}, NOrec~\cite{dalessandro2010norec} and UndoLog~\cite{felber2008dynamic} (with and without visible readers).

Both TL2 and NOrec follow the write-back design strategy and validate transactions at commit time. To enforce ACO in these implementations, transactions are allowed to enter the commit phase only when all transactions with lower age have been committed.
In order to aid the ordering for UndoLog, we exploit an age-based contention policy (i.e., always favor transactions with the lower age) to handle write-write conflicts.
In the \emph{visible readers} variant, the writer transaction aborts all active readers, while when readers are \emph{invisible} the writer retries multiple times if the object is locked, then it backs off.
STMLite uses a write-back implementation and replaces the need for constructing a read-set by leveraging signatures (Bloom Filters).
There is a tradeoff in determining the effective size of signatures, but the authors recommended a range of 32 to 1024. We used a signature of size 64 with the STL hashing function because it provided the best performance.
The number of threads in STMLite also includes its commit manager.

All competitors, including STMLite whose source code, to the best of our knowledge, is not publicly available, have been re-implemented atop the same baseline software framework so that all take advantage of the same low-level optimizations.
It is worth noting that competitors may provide different correctness guarantees (e.g., OWB provides TMS1 while NOrec/TL2 give opacity). %Additional implementation details have been provided in the supplementary material.

In our experiments, the ACO is defined in two ways. Unless otherwise specified, the index of the dominant for-loop that each benchmark uses to generate parallel code (e.g., transactions in STAMP) is used as transaction age. In some application with more complex patterns, such as nested loops, we inserted an atomic integer to define and assign ages.

Threads are pinned to cores. The policy is to use up all cores of one socket before moving to the other one.

We report the throughput for micro benchmarks and the application execution time for STAMP and some applications of PARSEC and SPEC200 benchmarks by varying the number of serving threads in the thread-pool (the datapoint at 1 thread shows the performance of the single-threaded transactional execution). We also compare our performance against the unordered algorithms, which do not use ACO. In this case, applications directly activate transactions on parallel because no ACO needs to be defined. Performance of the non-transactional single-threaded execution (green line) is also  included. 

We used two different machines for our experiments: micro benchmarks and STAMP have been evaluated on an AMD machine equipped with 2 Opteron 6168 CPUs, each with 12-core running at 1.9 GHz. The total memory available is 12 GB. Evaluation of PARSEC applications and SPEC2000 has been done using a Intel server hosting 4 Intel Xeon Platinum 8160. Results are the average of five runs.

\textbf{Micro Benchmark}. In our first set of experiments we consider the RSTM micro-benchmarks~\cite{SOFTWARESTM5} to evaluate the effect of different workload characteristics, such as the amount of operations per transaction, the transaction length, and the read/write ratio, on the performance. Each experiment included running half a million transactions. For each micro benchmark, we configured three types of transactions: short, long, and heavy. Both \emph{short} and \emph{heavy} have the same number of accesses (i.e., a random between 10 and 20), but the latter adds more local computation in between them (i.e., 100 CPU-ops). \emph{Long} transactions simply produce more transactional accesses (i.e., a random between 30 and 60).

Figure~\ref{fig:micro-all} summarizes the peak performance of all competitors. From that we can see the gap in performance between the ordered and unordered versions of the same algorithm: 26-56\% for TL2, 13-41\% for NOrec, 12-88\% for UL-vis, and 28-74\% for UL-invis.

As a general comment on these results, OUL and OUL-Steal outperform all other ordered versions of the algorithms. OUL-Steal excels for write loads and performs equally to OUL in read loads; OWB outperforms all write-back based implementations in most benchmarks.
At high thread count, STMLite suffers from false conflicts due to the use of signatures. However, at low number of threads (less than 8) and with \emph{Long} transactions it achieves a higher peak throughput than Ordered TL2 and Ordered NOrec, because it benefits from the quick validation using signatures.
For the UL-inv algorithm, we found that the readers' visibility was crucial; without this information, the algorithm may abort a lower age transaction (using timeout) while some higher age transaction holds the read shared lock.
On the other hand, these higher age transactions cannot commit before their order comes, hence they timeout.

\begin{figure}[h]
	\begin{subfigure}[b]{0.23\textwidth} \includegraphics[trim=1cm 3cm 0cm 0cm,clip=true,scale=0.7]{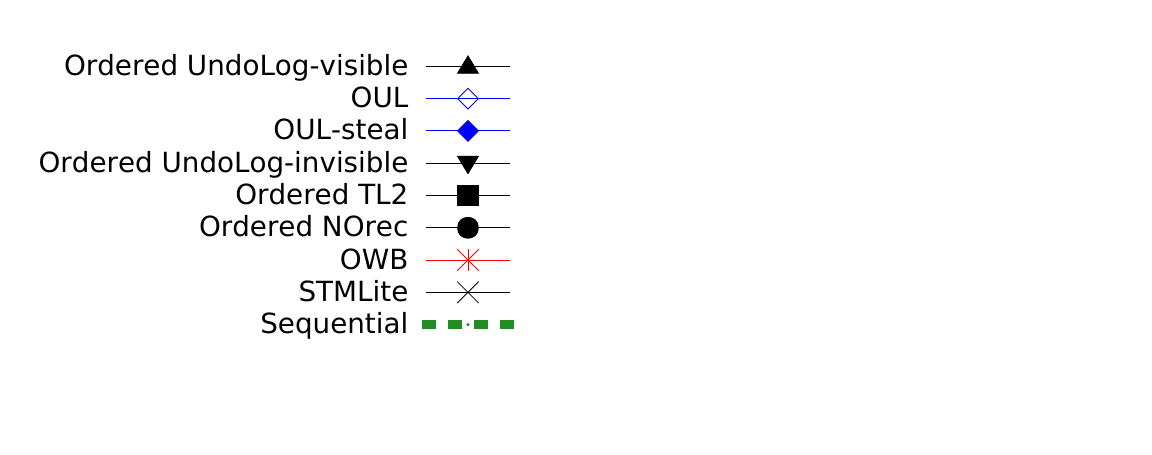}  \end{subfigure}
	\begin{subfigure}[b]{0.23\textwidth} \includegraphics[scale=0.35]{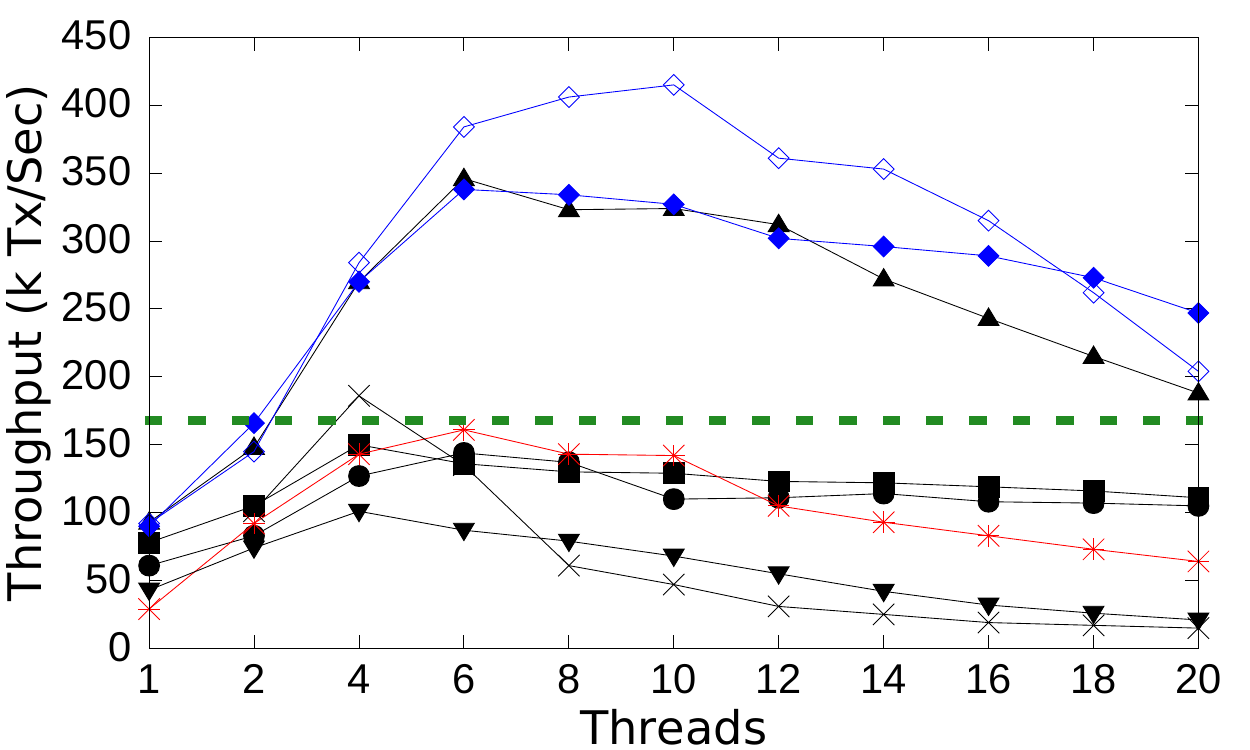} \vspace{-1.6em} \caption{Disjoint-Long} \label{fig:disjoint-long-speedup} \end{subfigure}
    \begin{subfigure}[b]{0.23\textwidth} \includegraphics[scale=0.35]{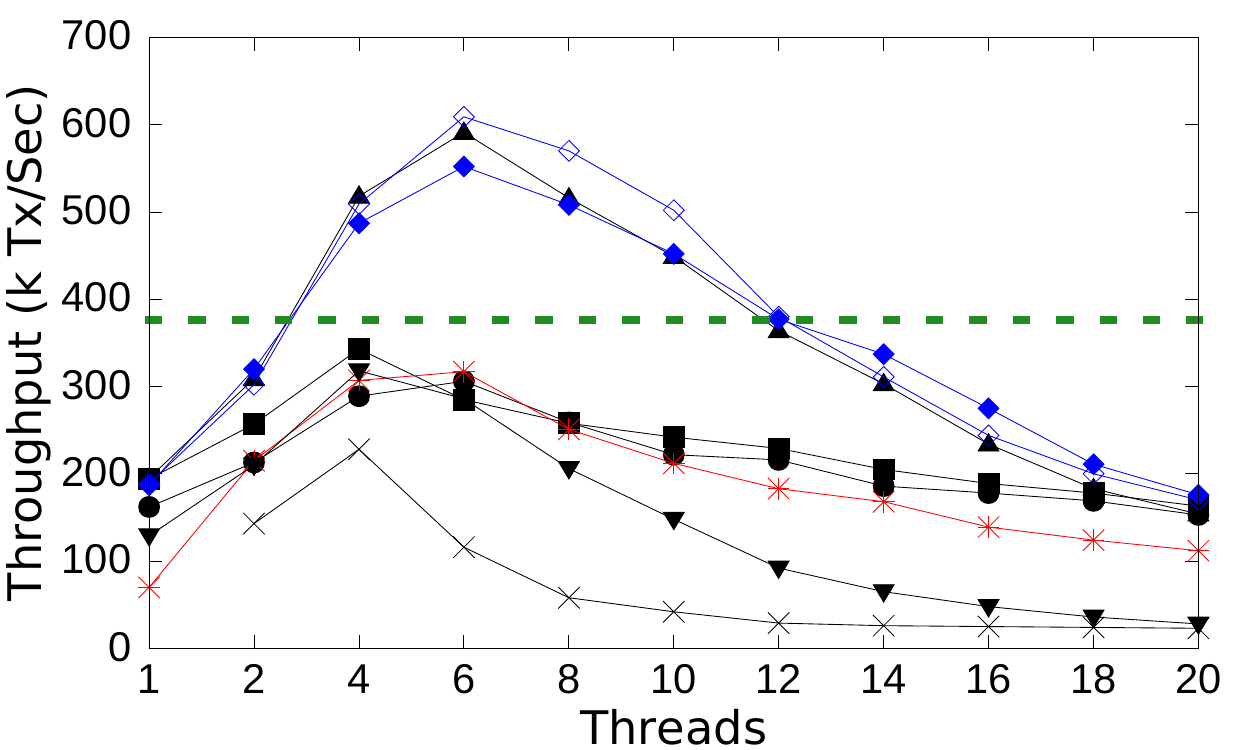} \vspace{-1.6em} \caption{Disjoint-Short} \label{fig:disjoint-short-speedup} \end{subfigure}
    \hspace{3pt}
    \begin{subfigure}[b]{0.23\textwidth} \includegraphics[scale=0.35]{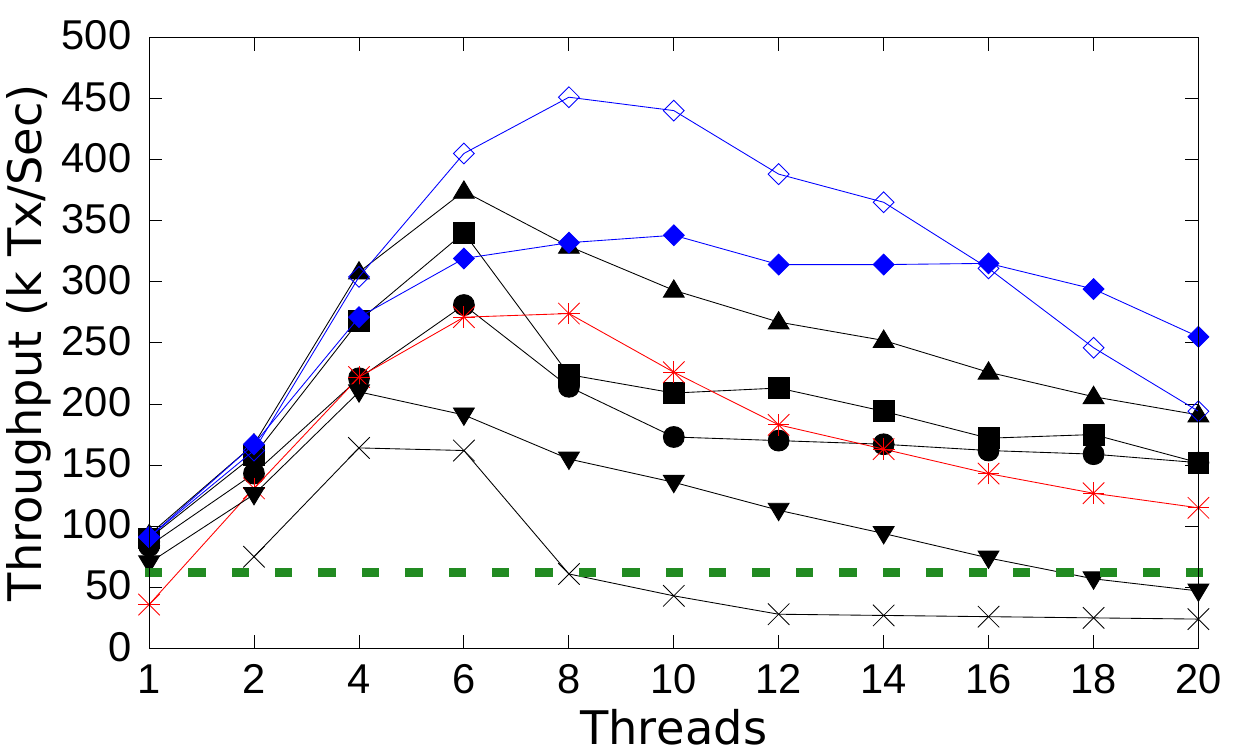} \vspace{-1.6em} \caption{Disjoint-Heavy} \label{fig:disjoint-heavy-speedup} \end{subfigure}
	\begin{subfigure}[b]{0.23\textwidth} \includegraphics[scale=0.35]{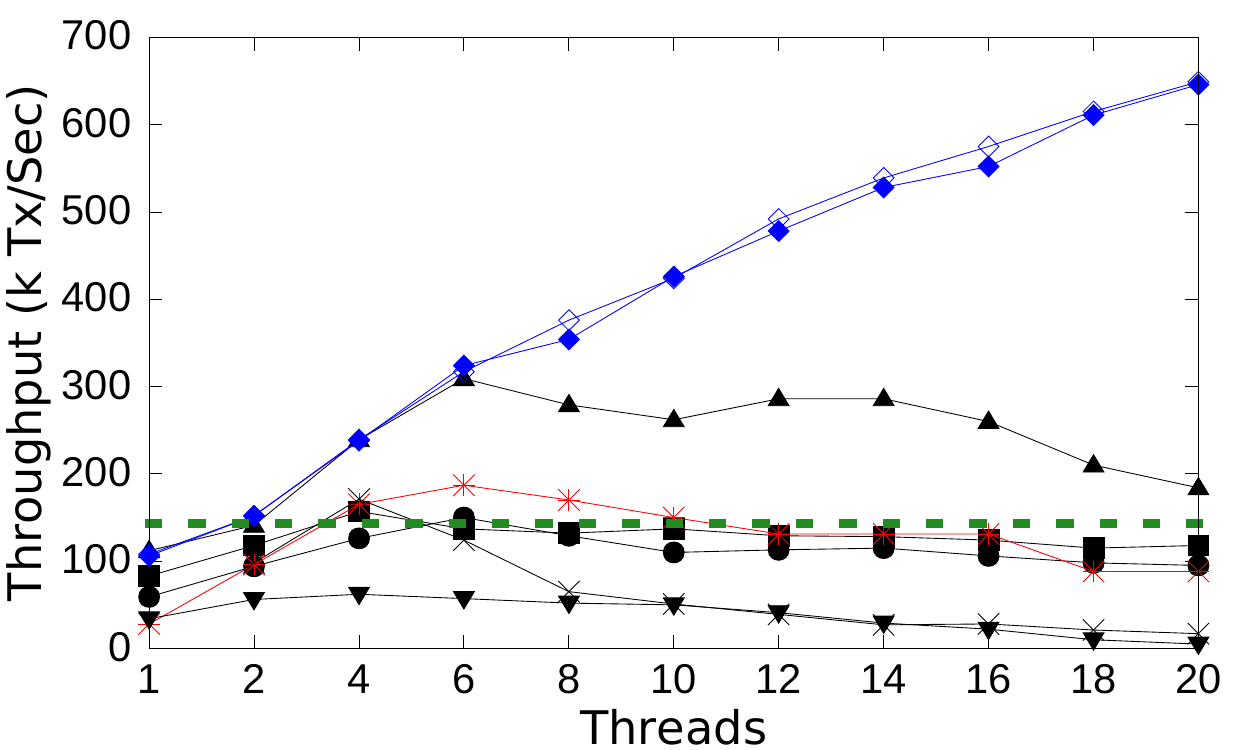} \vspace{-1.6em} \caption{RNW1-Long} \label{fig:rnw1-long-speedup} \end{subfigure}
    \hspace{3pt}
    \begin{subfigure}[b]{0.23\textwidth} \includegraphics[scale=0.35]{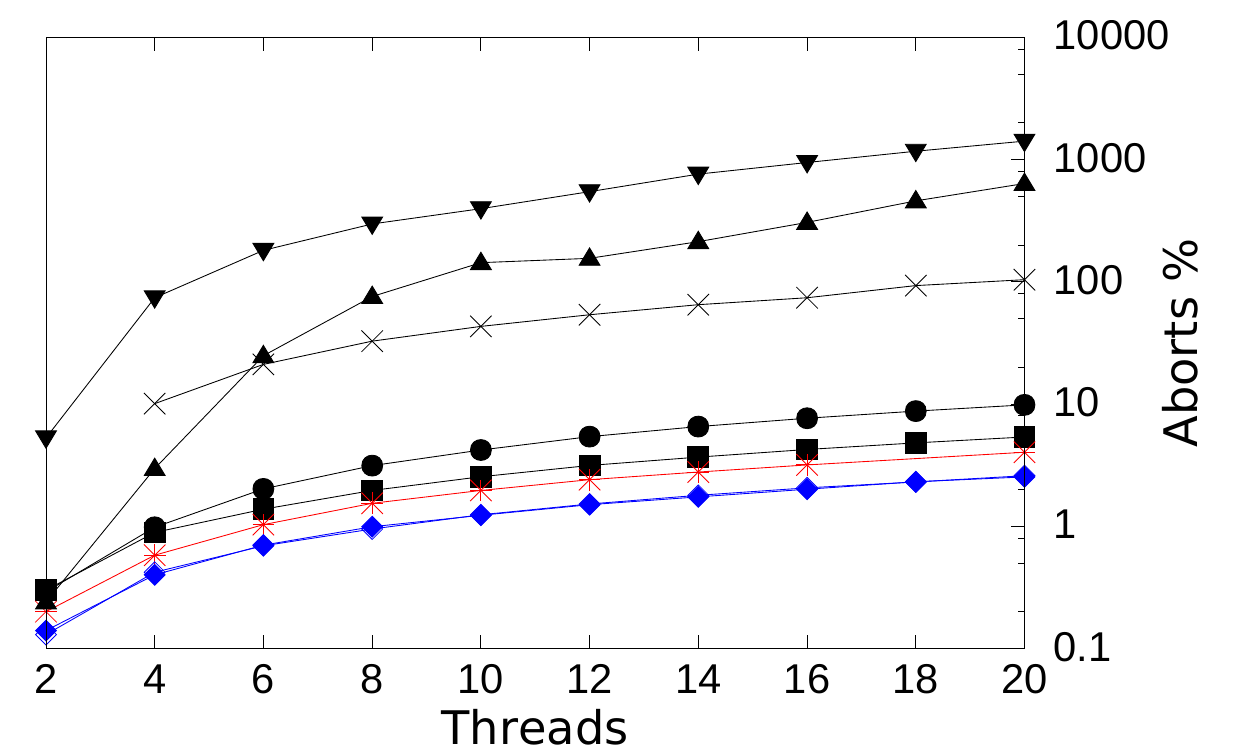} \vspace{-1.6em} \caption{RNW1-Long Aborts} \label{fig:rnw1-long-aborts} \end{subfigure}
    \begin{subfigure}[b]{0.23\textwidth} \includegraphics[scale=0.35]{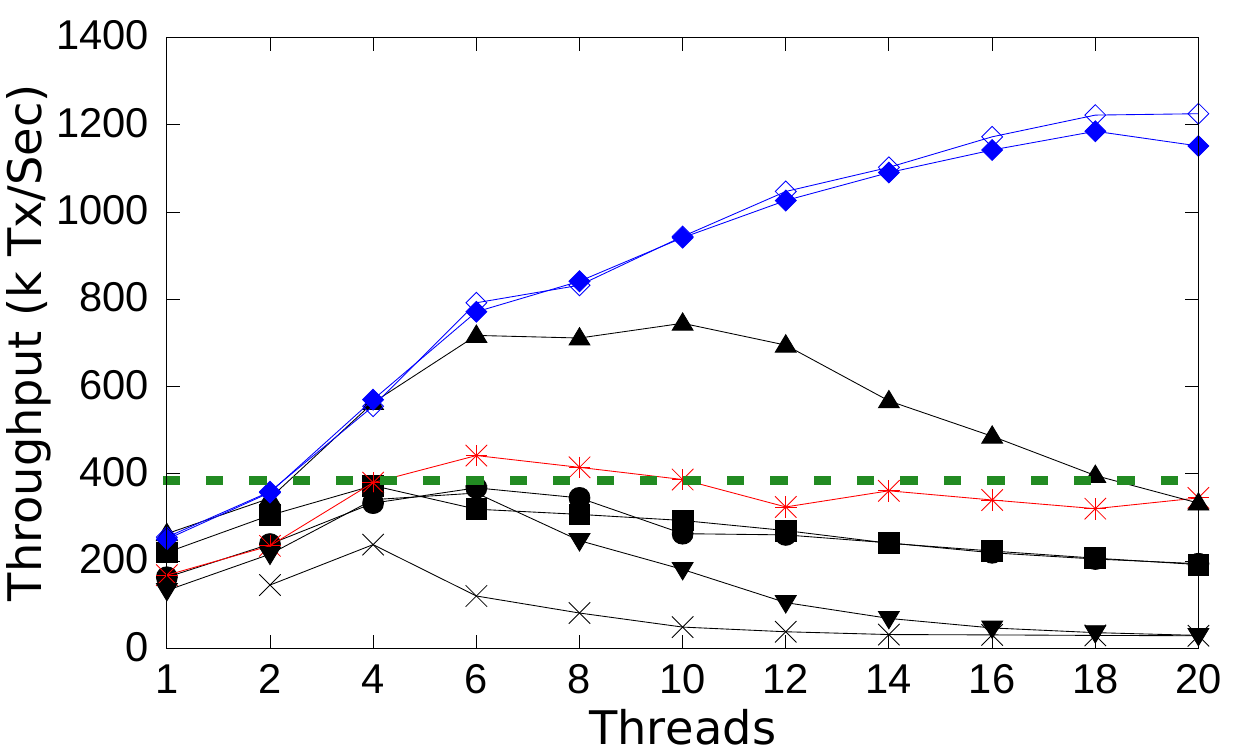} \vspace{-1.6em} \caption{RNW1-Short} \label{fig:rnw1-short-speedup} \end{subfigure}
    \hspace{3pt}
    \begin{subfigure}[b]{0.23\textwidth} \includegraphics[scale=0.35]{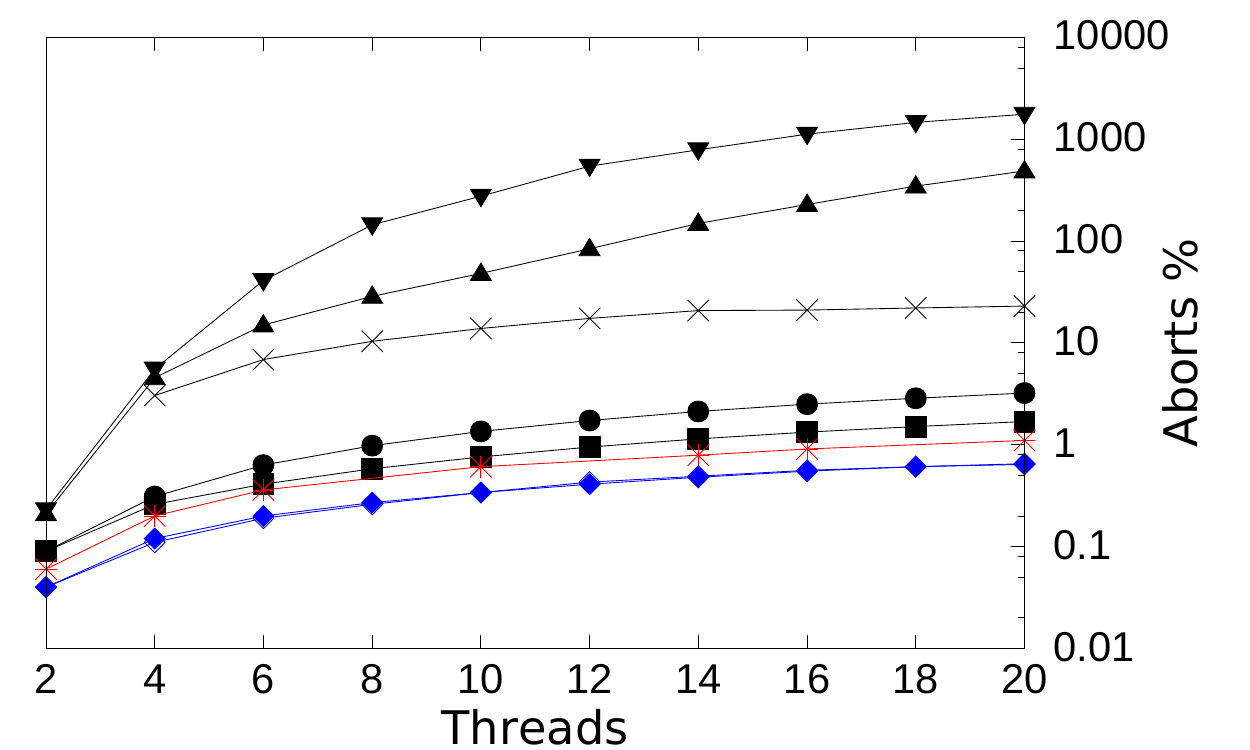} \vspace{-1.6em} \caption{RNW1-Short Aborts} \label{fig:rnw1-short-aborts} \end{subfigure}
    \begin{subfigure}[b]{0.23\textwidth} \includegraphics[scale=0.35]{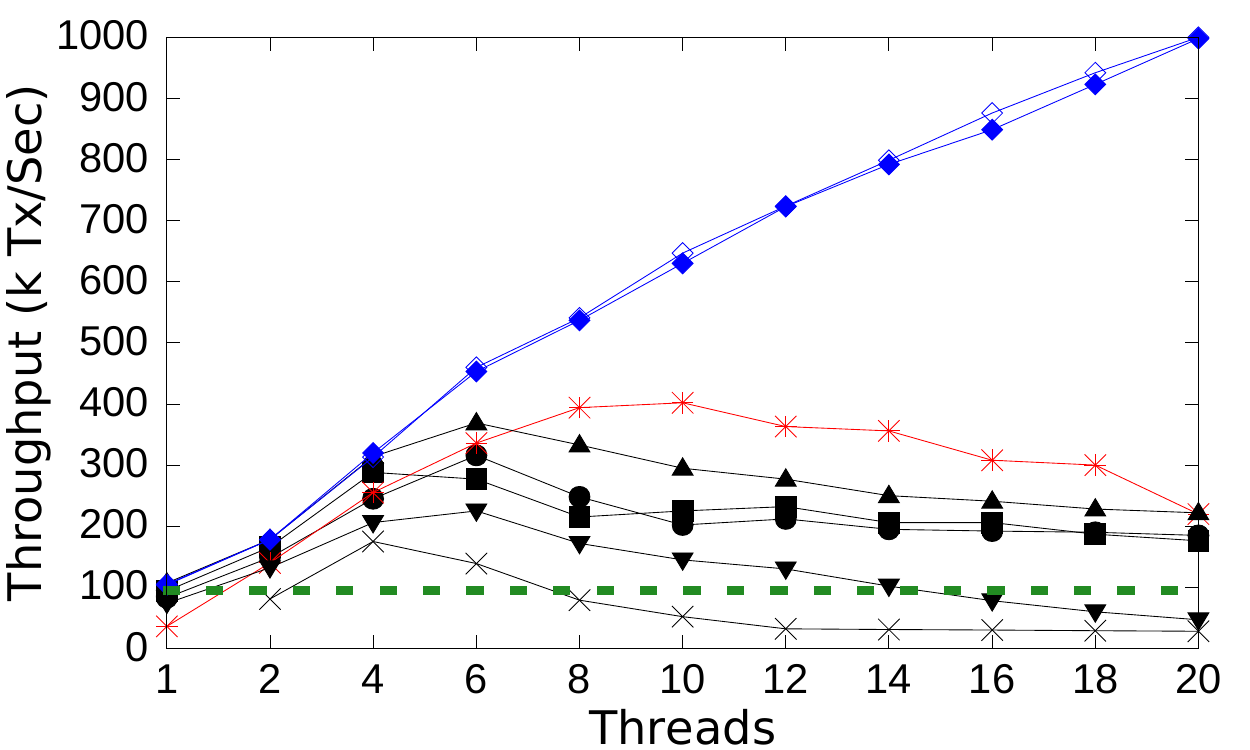} \vspace{-1.6em} \caption{RNW1-Heavy} \label{fig:rnw1-heavy-speedup} \end{subfigure}
    \hspace{3pt}
    \begin{subfigure}[b]{0.23\textwidth} \includegraphics[scale=0.35]{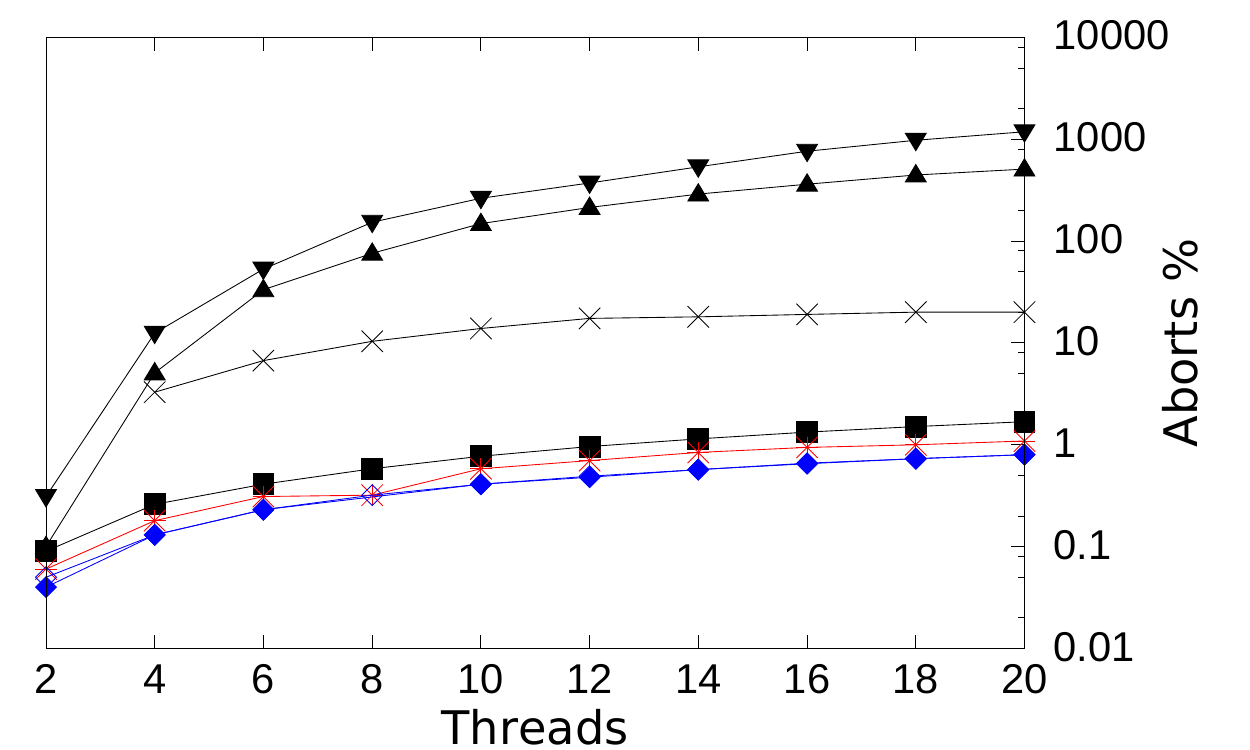} \vspace{-1.6em} \caption{RNW1-Heavy  Aborts} \label{fig:rnw1-heavy-aborts} \end{subfigure}
    \vspace{-15pt}
    \caption{Disjoint \ref{fig:disjoint-long-speedup} -- \ref{fig:disjoint-heavy-speedup} and ReadNWrite1 \ref{fig:rnw1-long-speedup} -- \ref{fig:rnw1-heavy-aborts}.}
    \label{fig:rnw1}
%    \vspace{-10pt}
\end{figure}

In configurations where the performance of the sequential (non-transactional) execution is faster than many ordered algorithms, our solutions outperform it, letting parallelism pay off. However, there are two benchmarks with long transactions where the sequential execution is faster. These workloads represent unfavorable scenarios for processing ordered transactions because of the high cost of aborting transactions (possible repeatedly) due to ACO violation.

The \textit{DisjointBench} (Figures~\ref{fig:disjoint-long-speedup}-\ref{fig:disjoint-heavy-speedup}) produces a workload with no conflict between concurrent transactions. Every transaction accesses a different set of addresses with read and write operations.
In all configurations, OUL achieves the best throughput, while OUL-Steal suffers from the overhead of its lock management scheme without actually gaining from that, as the disjoint transactions do not have any shared accesses. UL-vis achieves a throughput near to OUL-Steal, thanks to the simplicity of its immediate write strategy. In all the three plots is visible a peak performance around 6 threads. This shape is the consequence of NUMA latency~\cite{DBLP:conf/spaa/BrownKLL16,DBLP:conf/wdag/DalyHSP18}, which can be appreciated in this configuration more than in others due to the absence of data contention.

\begin{figure}[h]
	\begin{subfigure}[b]{0.23\textwidth} \includegraphics[scale=0.35]{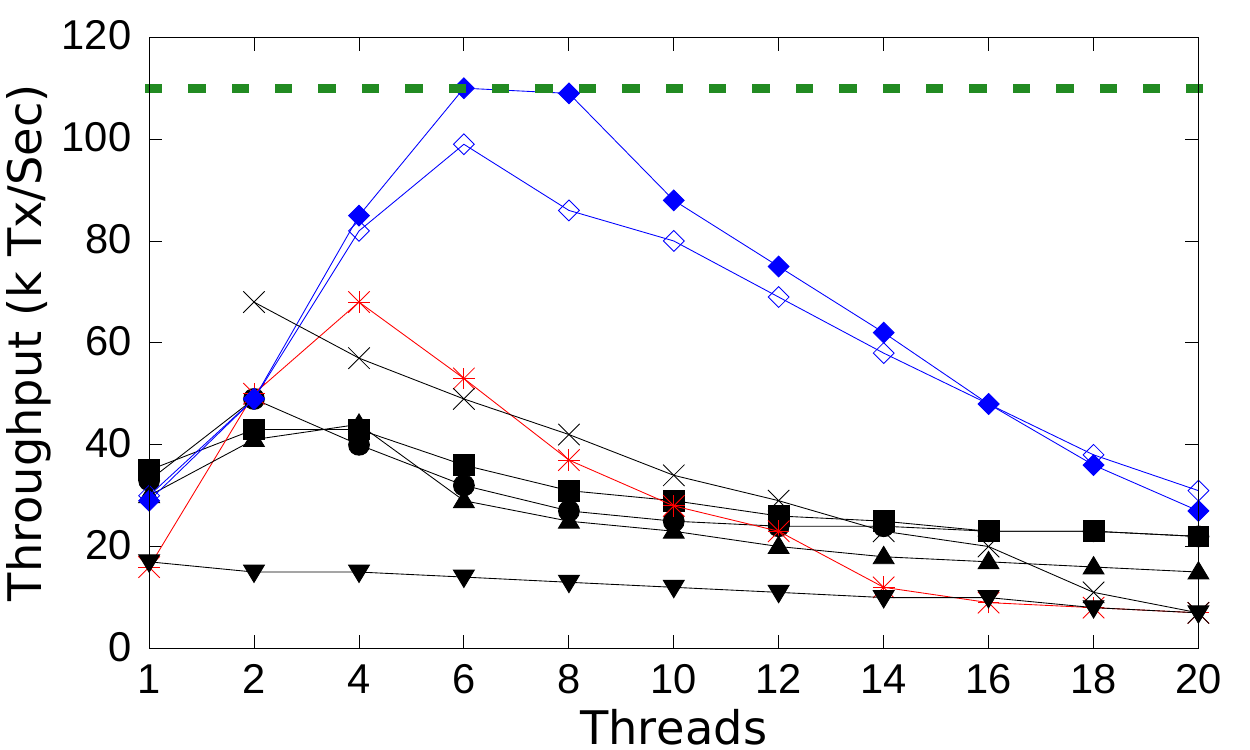} 	\vspace{-1.5em}	\caption{RWN-Long} \label{fig:rwn-long-speedup} \end{subfigure}
	    \hspace{3pt}
    \begin{subfigure}[b]{0.23\textwidth} \includegraphics[scale=0.35]{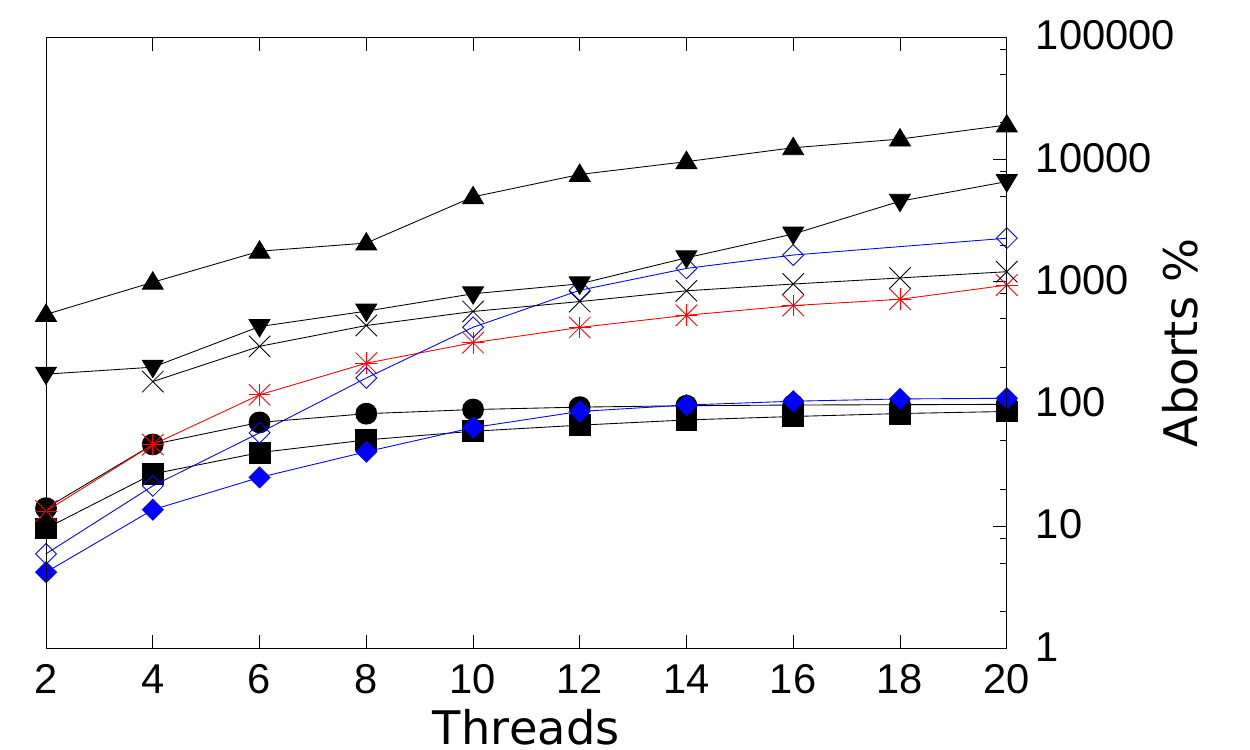}  	\vspace{-1.5em}	\caption{RWN-Long Aborts} \label{fig:rwn-long-aborts} \end{subfigure}
    \begin{subfigure}[b]{0.23\textwidth} \includegraphics[scale=0.35]{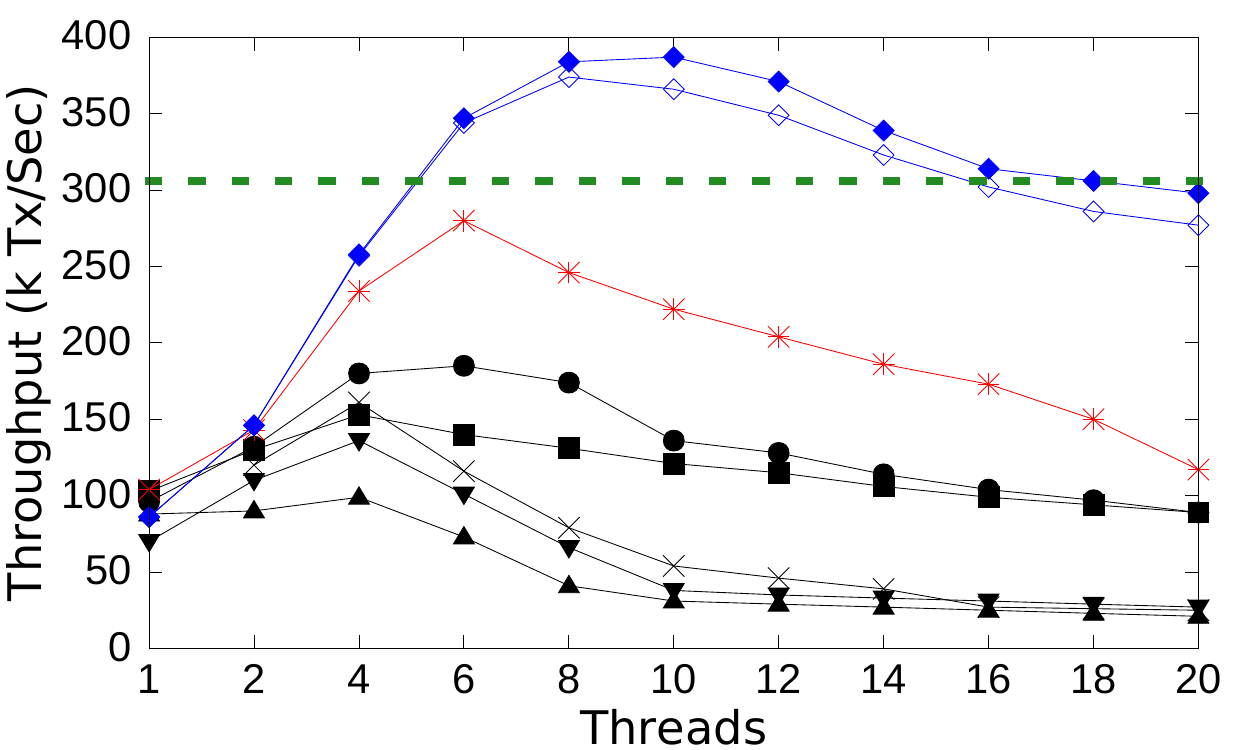} 	\vspace{-1.5em}	\caption{RWN-Short} \label{fig:rwn-short-speedup} \end{subfigure}
        \hspace{3pt}
    \begin{subfigure}[b]{0.23\textwidth} \includegraphics[scale=0.35]{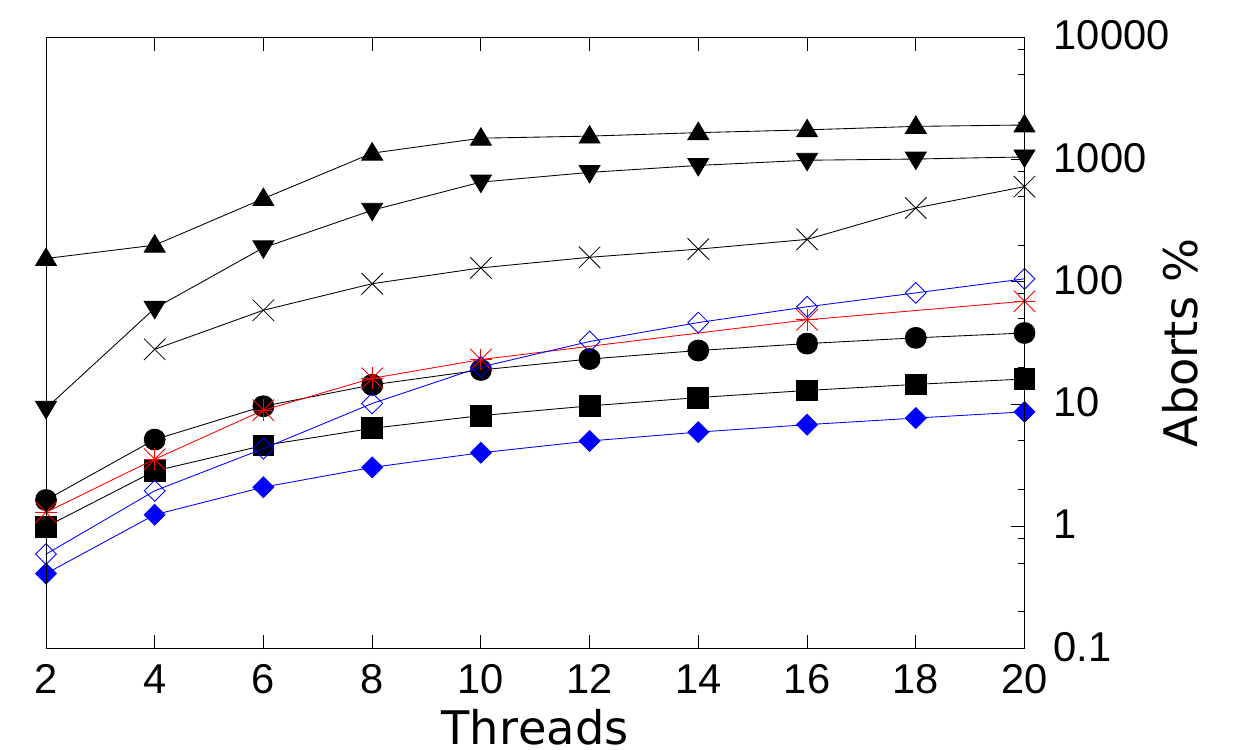} 	\vspace{-1.5em} \caption{RWN-Short Aborts} \label{fig:rwn-short-aborts} \end{subfigure}
    \begin{subfigure}[b]{0.23\textwidth} \includegraphics[scale=0.35]{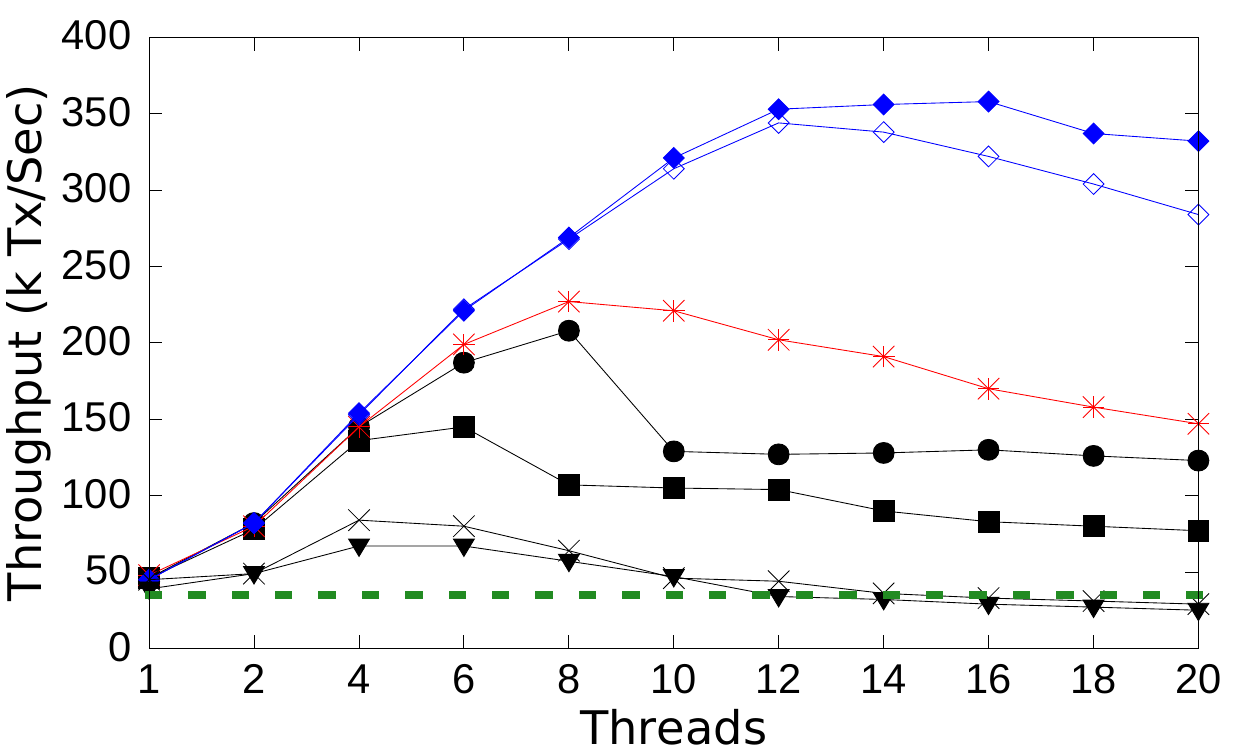} 	\vspace{-1.5em}	\caption{RWN-Heavy} \label{fig:rwn-heavy-speedup} \end{subfigure}
        \hspace{3pt}
    \begin{subfigure}[b]{0.23\textwidth} \includegraphics[scale=0.35]{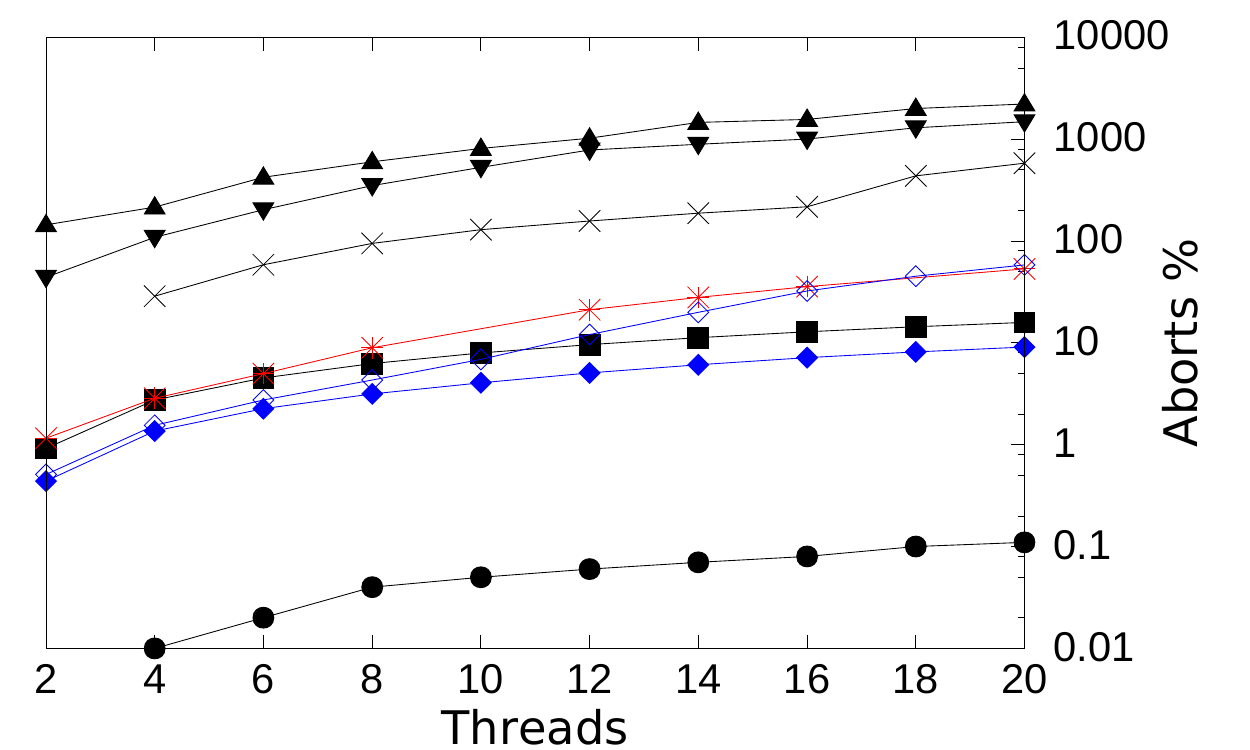} 	\vspace{-1.5em}	\caption{RWN-Heavy Aborts} \label{fig:rwn-heavy-aborts} \end{subfigure}
    \begin{subfigure}[b]{0.23\textwidth} \includegraphics[scale=0.35]{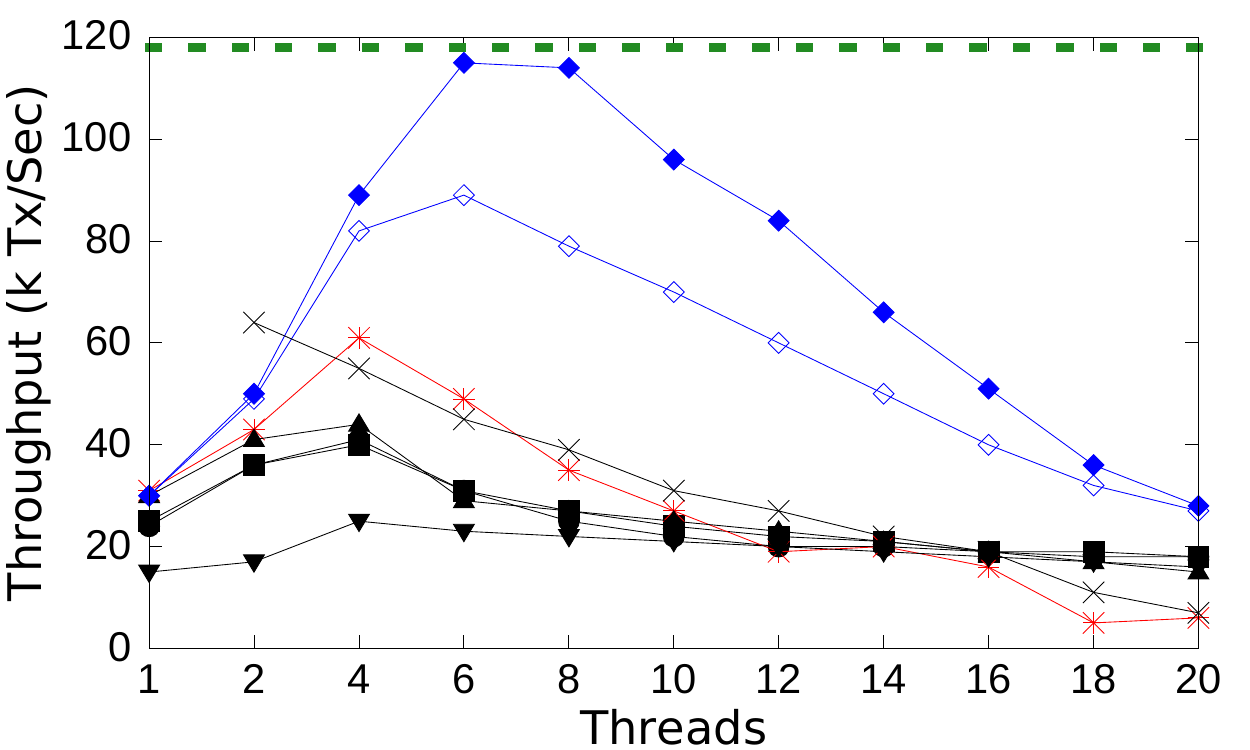} 	\vspace{-1.5em}	\caption{MCAS-Long} \label{fig:mcas-long-speedup} \end{subfigure}
    \hspace{3pt}
    \begin{subfigure}[b]{0.23\textwidth} \includegraphics[scale=0.35]{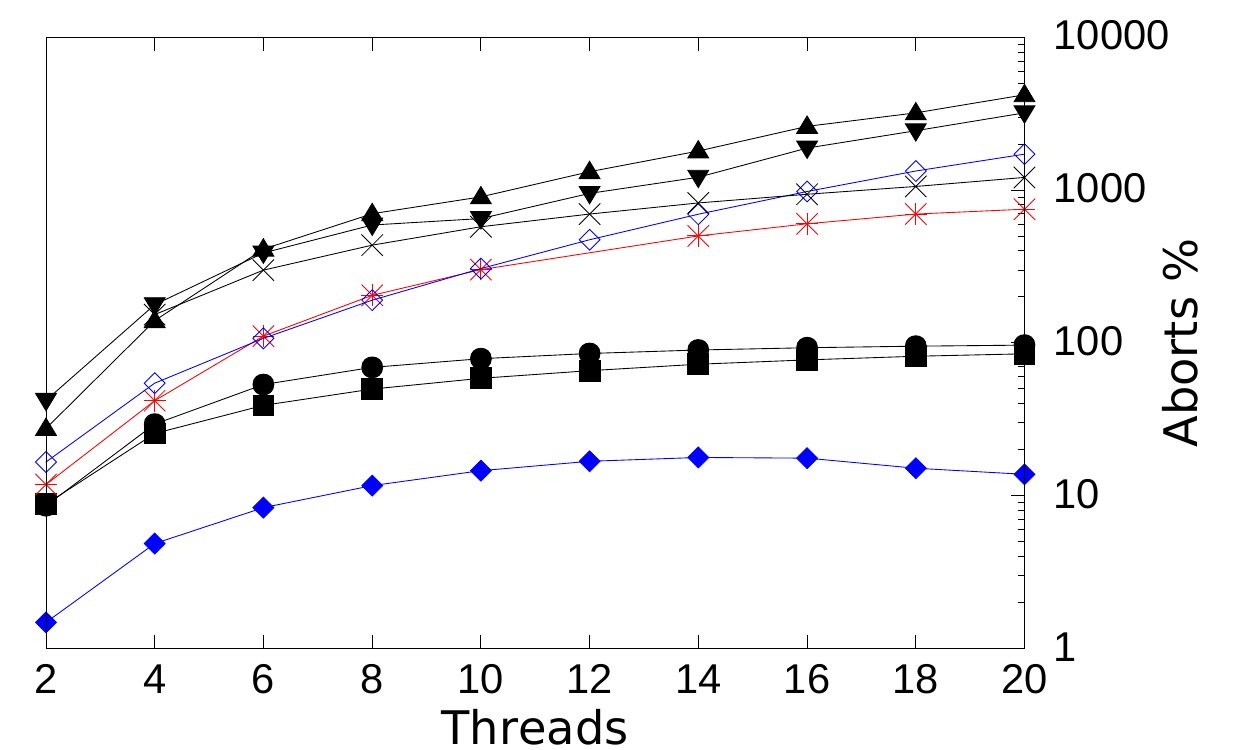} 	\vspace{-1.5em}	\caption{MCAS-Long Aborts} \label{fig:mcas-long-aborts} \end{subfigure}
    \begin{subfigure}[b]{0.23\textwidth} \includegraphics[scale=0.35]{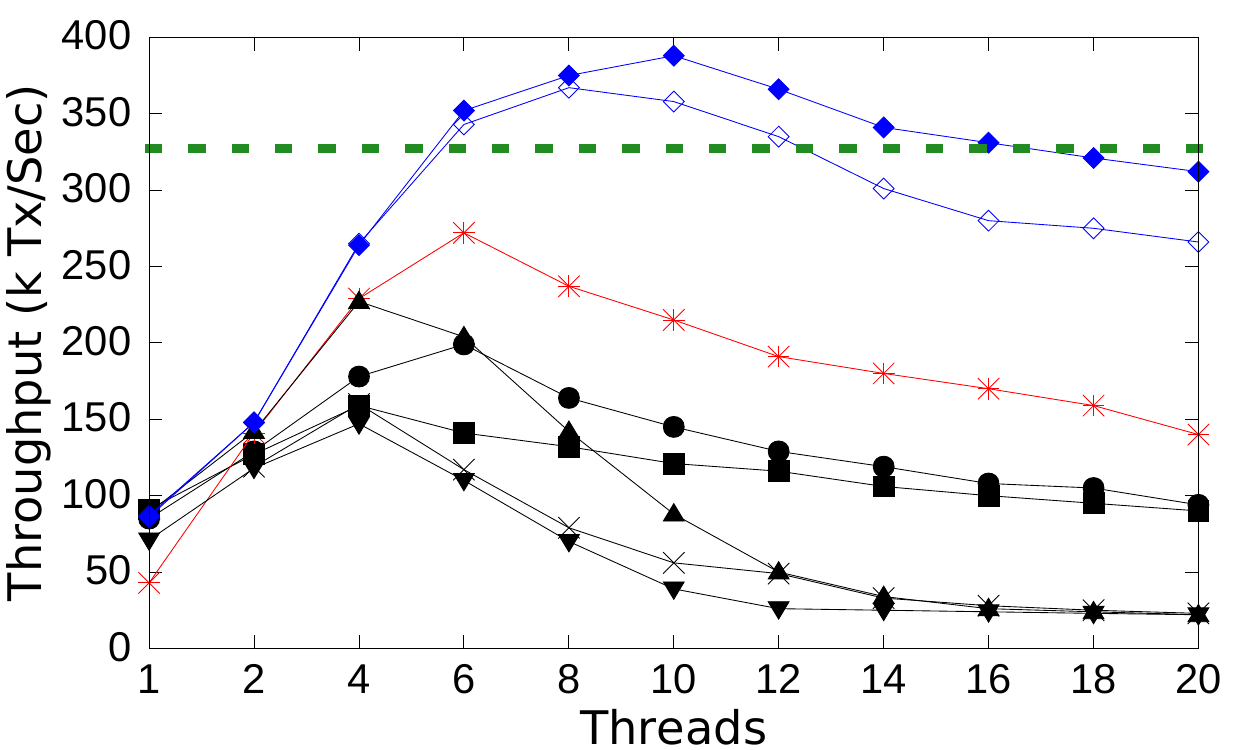} \vspace{-1.5em}	\caption{MCAS-Short} \label{fig:mcas-short-speedup} \end{subfigure}
    \hspace{3pt}
    \begin{subfigure}[b]{0.23\textwidth} \includegraphics[scale=0.35]{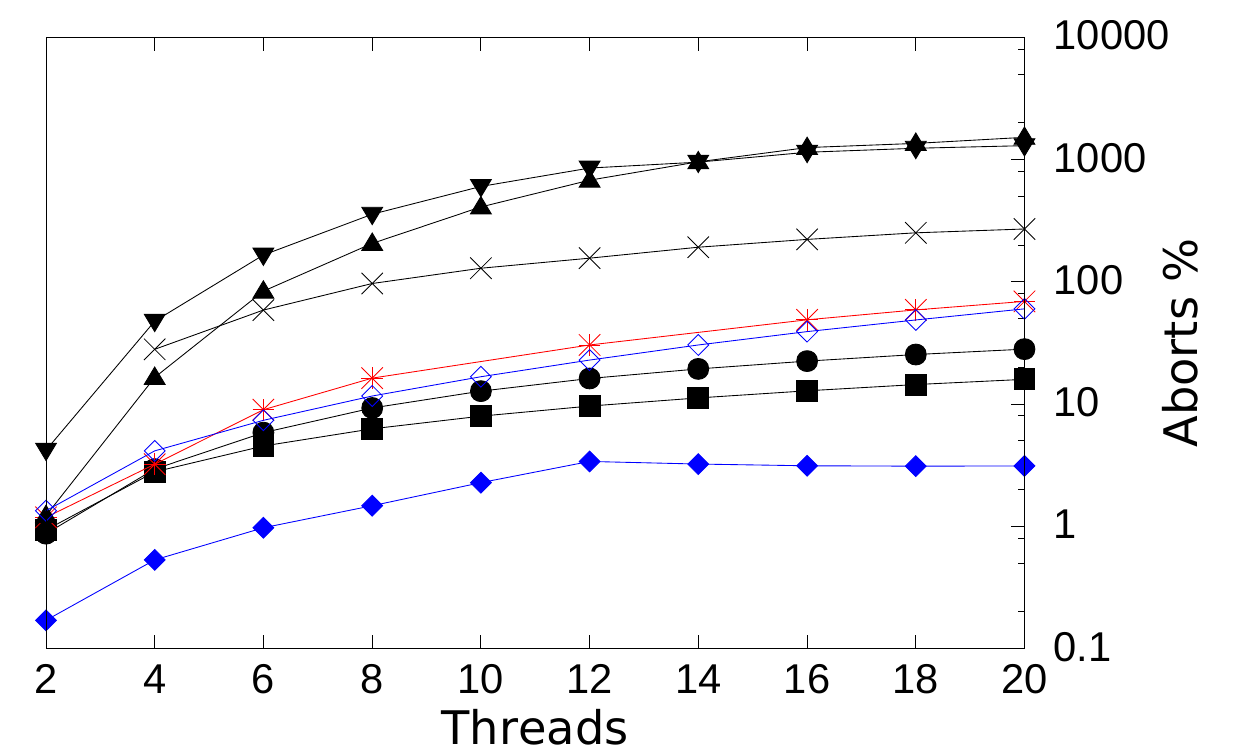} 	\vspace{-1.5em}	\caption{MCAS-Short  Aborts} \label{fig:mcas-short-aborts} \end{subfigure}
    \begin{subfigure}[b]{0.23\textwidth} \includegraphics[scale=0.35]{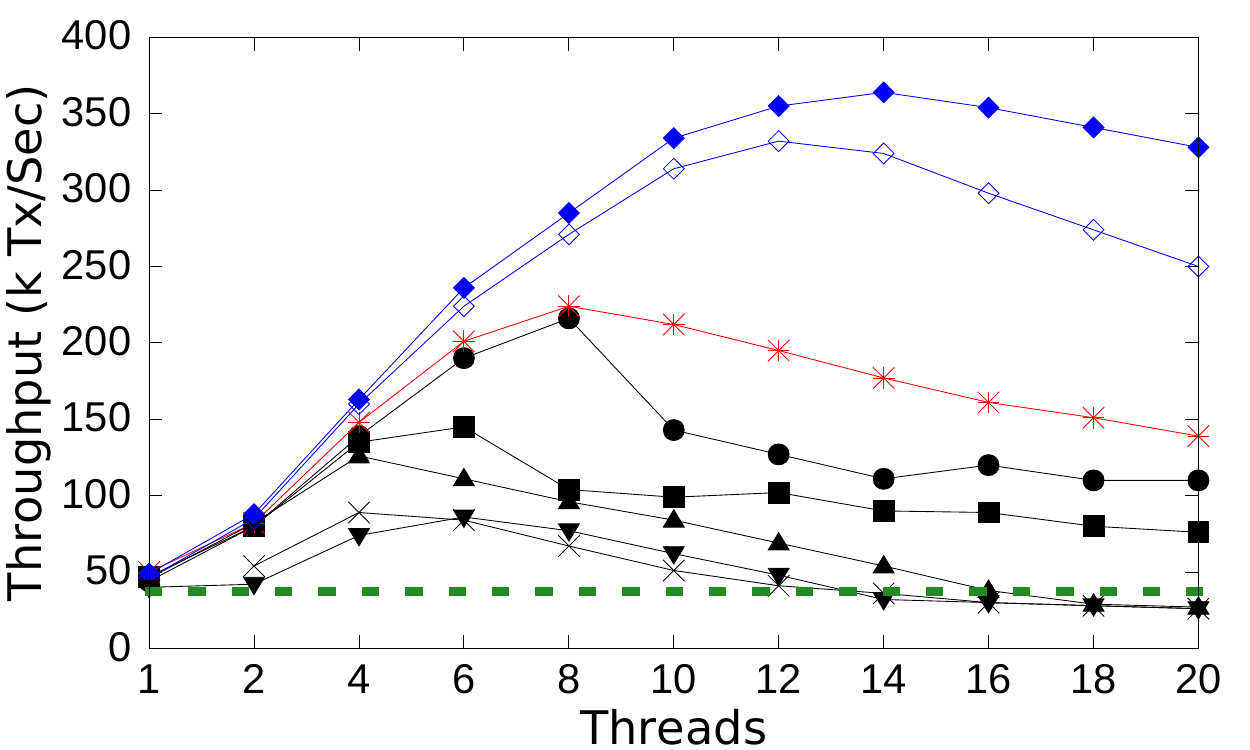} \vspace{-1.5em}	\caption{MCAS-Heavy} \label{fig:mcas-heavy-speedup} \end{subfigure}
    \hspace{3pt}
    \begin{subfigure}[b]{0.23\textwidth} \includegraphics[scale=0.35]{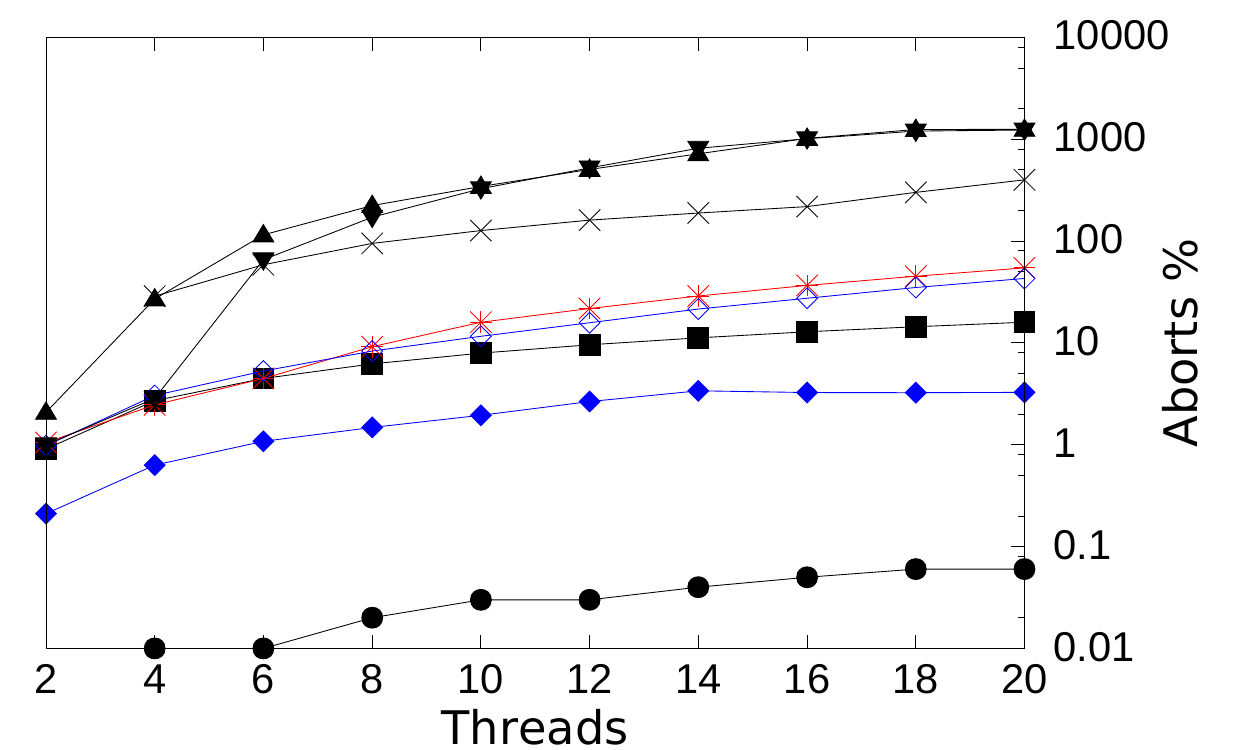} 	\vspace{-1.5em}	\caption{MCAS-Heavy Aborts} \label{fig:mcas-heavy-aborts} \end{subfigure}
   \vspace{-16pt}
    \caption{RWN \ref{fig:rwn-long-speedup} -- \ref{fig:rwn-heavy-aborts}; MCAS \ref{fig:mcas-long-speedup} -- \ref{fig:mcas-heavy-aborts}}
    \label{fig:mcas}
\end{figure}

In addition to that, without aborts we can show the transactional access overhead for each of all competitors. It is intuitive that UndoLog algorithms (UL-vis, UL-inv, OUL, OUL-Steal) benefit from having the values already in memory, thus they outperform others. In fact, the UndoLog's main drawback is the costly abort, which never happens in this benchmark. With Long transactions (Figure~\ref{fig:disjoint-long-speedup}), STMLite benefits from eliminating lock usage at the write-back phase and it has minimal overhead at low numbers of threads. On the other hand, for Short transactions (Figures~\ref{fig:disjoint-short-speedup} and~\ref{fig:disjoint-heavy-speedup}) the Ordered TL2 algorithm performs better. OWB has a moderate overhead relative to the other write-back algorithms.

In \textit{ReadNWrite1Bench} (Figures~\ref{fig:rnw1-long-speedup}-\ref{fig:rnw1-heavy-aborts}), the transaction reads N locations and writes one.
Since transaction write-set is very small, the number of aborts is low. Similarly, UndoLog algorithms excel here as well.
With \emph{long} and \emph{heavy} transactions (Figure~\ref{fig:rnw1-long-speedup},~\ref{fig:rnw1-heavy-speedup}), the processing done by workers is overweights the overhead due to single thread transaction validator, so both OUL and OUL-Steal scales well with increasing the number of workers. On the other hand, the validator represents a performance bottleneck for short transactions (Figure \ref{fig:rnw1-short-speedup}), resulting in a slightly lower scalability.

In \textit{ReadWriteN} (Figures~\ref{fig:rwn-long-speedup}-\ref{fig:rwn-heavy-aborts}), each transaction reads N locations, and then writes to other N locations. The large transaction write-set introduces a challenge for both undo-log (increases the number of aborts) and write-buffer algorithms (delay at commit time). 
The cooperative execution enables OUL, OUL-Steal and OWB to outperforms all other algorithms at all workloads.
OUL-Steal outperforms OUL by 10\% because it significantly reduces the number of aborts (Figures~\ref{fig:rwn-long-aborts},~\ref{fig:rwn-short-aborts}, and~\ref{fig:rwn-heavy-aborts}. 

\begin{figure}[h]
\begin{subfigure}[b]{0.5\textwidth} \centering \includegraphics[trim=0cm 4.3cm 0cm 0cm,clip=true,scale=0.5]{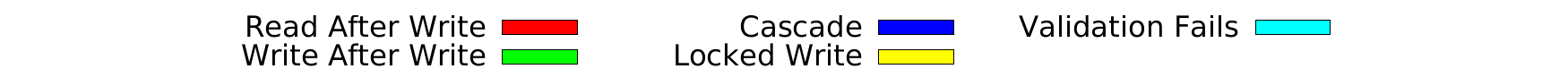}
\end{subfigure}
    \begin{subfigure}[b]{0.23\textwidth} \centering \includegraphics[scale=0.45]{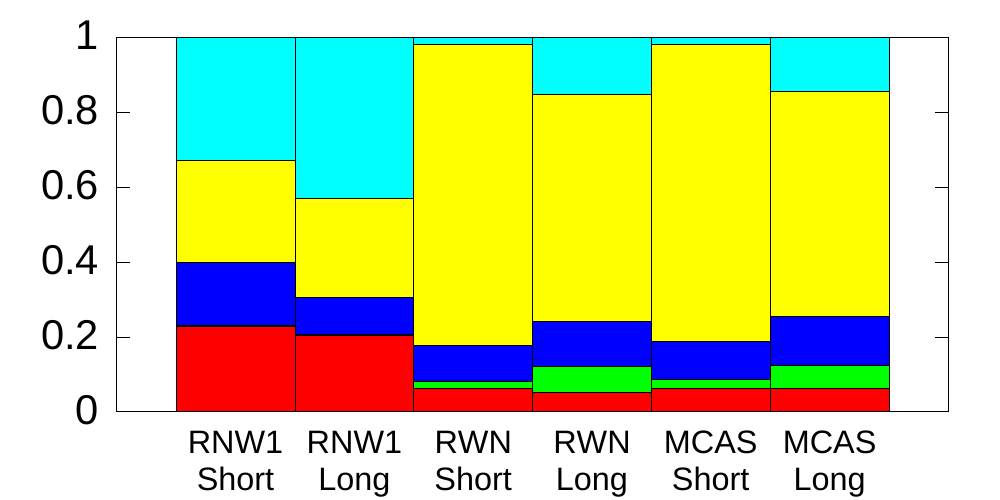} \vspace{-1.5em} \caption{OWB Algorithm} \label{fig:owb-aborts} \end{subfigure}
    \begin{subfigure}[b]{0.23\textwidth} \centering \includegraphics[scale=0.45]{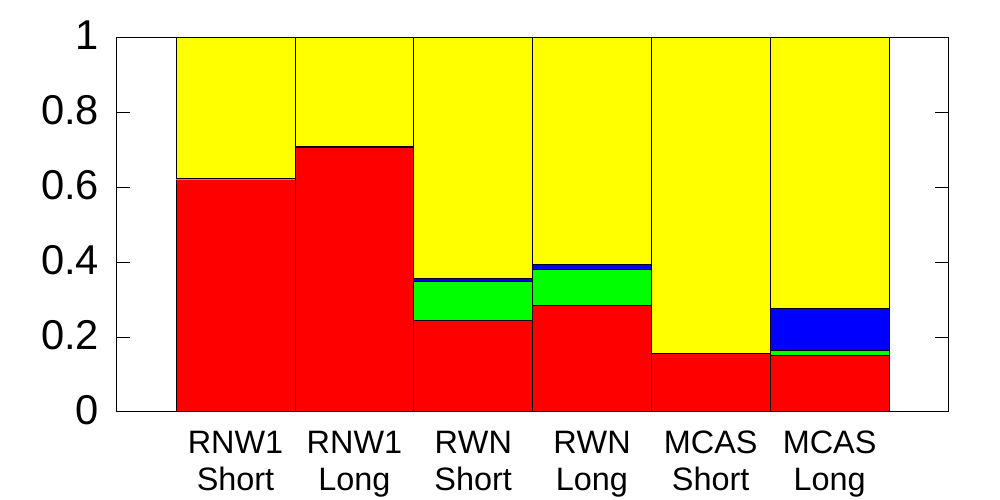} \vspace{-1.5em} \caption{OUL Algorithm} \label{fig:oul-aborts}  \end{subfigure}
    \begin{subfigure}[b]{0.23\textwidth} \centering \includegraphics[scale=0.45]{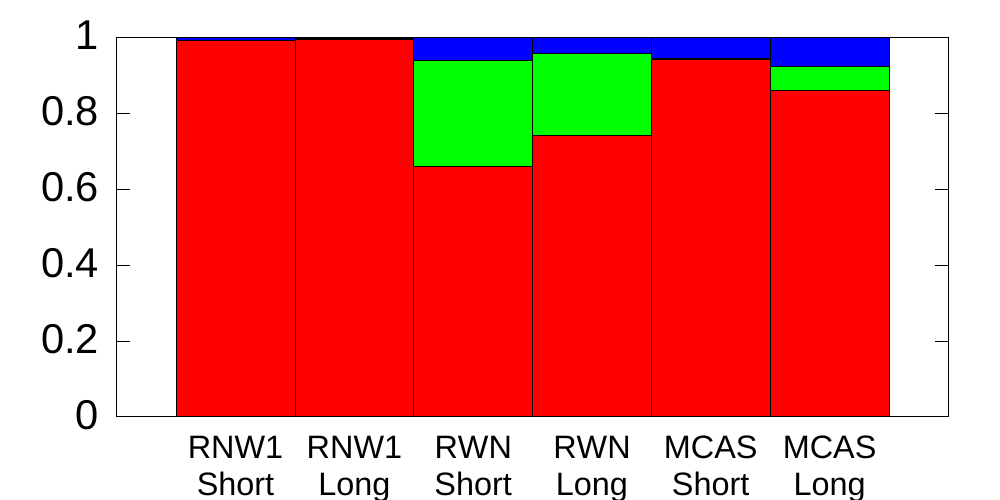} \vspace{-1.5em} \caption{OUL-Steal Algorithm} \label{fig:oul-steal-aborts} \end{subfigure}
    \begin{subfigure}[b]{0.24\textwidth} \centering \includegraphics[scale=0.45]{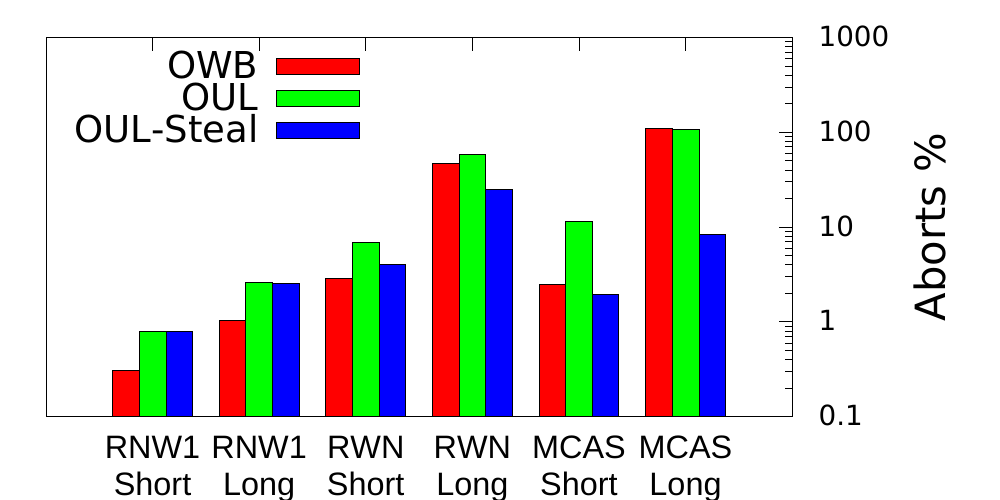} \vspace{-1.5em} \caption{Aborts Percentage} \label{fig:all-aborts} \end{subfigure}
    \vspace{-1.5em}
    \caption{Aborts Breakdown}
    \label{fig:aborts}
\end{figure}

\textit{MCASBench} performs a multi-word compare-and-swap, by reading and writing N consecutive locations. Similar to ReadWriteN, the write-set is large but the abort probability is lower than before because each pair of read/write acts on the same location.
Figures~\ref{fig:mcas-long-speedup}-\ref{fig:mcas-heavy-aborts} illustrate the impact of increasing workers with the different workloads. We noticed a similar trend to \emph{ReadWriteN}.

The breakdown of the abort reasons for OWB, OUL, and OUL-Steal is shown in Figure~\ref{fig:aborts}. Aborts are measured for the number of workers that achieved the maximum throughput.

In OWB (Figure~\ref{fig:owb-aborts}), with \emph{RNW1bench} aborts due to validation failure represent the main reason; while in write-intensive benchmarks, such as \emph{RWNbench} and \emph{MCASbench}, aborts are mainly ($65\%$-$82\%$) due to concurrent commits (\emph{Locked Write}). However, only $3\%$ of these cases falls in WAW, which means that OWB can benefit from the lock-steal optimization and save a considerable amount of aborts. 
However, applying lock-steal on OWB would complicate the design and the validation procedure. The reason is that transactions use commit-time locking and rely on the version number to validate their read-set. With lock-steal, multiple writers would increment the version number, thereby readers would not be able to do the validation simply.

For OUL and write-intensive benchmarks, concurrent writes generate between 70\% to 85\% of total aborts;
a WAW represents at most 10\% of them. In OUL-Steal, stealing the lock eliminates the problem of concurrent writes, and narrows write-write conflicts to only the WAW anomaly. However, it introduces several changes to the abort characteristics: a writer transaction that steals the lock becomes able to abort any invalid speculative readers earlier than before. This was reflected on increasing the number of \emph{Read After Write} aborts; the probability of triggering cascading aborts is increased if compared to OUL (Figures~\ref{fig:oul-aborts},~\ref{fig:oul-steal-aborts}); and the total number of aborts of OUL is reduced by one order of magnitude (Figures~\ref{fig:rnw1-long-aborts},~\ref{fig:rnw1-short-aborts},~\ref{fig:rwn-long-aborts},~\ref{fig:rwn-short-aborts},~\ref{fig:mcas-long-aborts},~\ref{fig:mcas-short-aborts},~\ref{fig:all-aborts}).

Although OUL-Steal substantially reduces the number of aborts, the speed-up is on average 20\%. The reasons for that are: the abort procedure for OUL-Steal is longer than OUL (2-4$\times$ in our experiments) because it involves recursive rollback for stolen locks.
This outweighs the reduction of the number of aborts; and OUL uses encounter time locking, thus aborts are detected at an early stage. This reduces the impact of aborting. In contrast, lazy algorithms (e.g., OWB) are greatly affected by aborts because the whole transaction needs to be re-executed since the invalidation is detected at commit time.
It is worth noting that, in OUL algorithms the abort cost differs according to the transaction type. In fact, aborting a write transaction requires restoring its original value, thus forcing the other transaction involved in the conflict to wait for the restoration of old written values; whereas aborting the readers is cheaper.

Figure~\ref{fig:all-aborts} shows the number of aborts in the maximum throughput scenario. OUL experiences more aborts than OWB because of the eager accesses; OUL-Steal avoids this drawback and experiences lesser, yet longer, aborts.

\begin{figure}
\begin{subfigure}[b]{0.5\textwidth} \centering \includegraphics[trim=2cm 3.5cm 0cm 0cm,clip=true,scale=0.5]{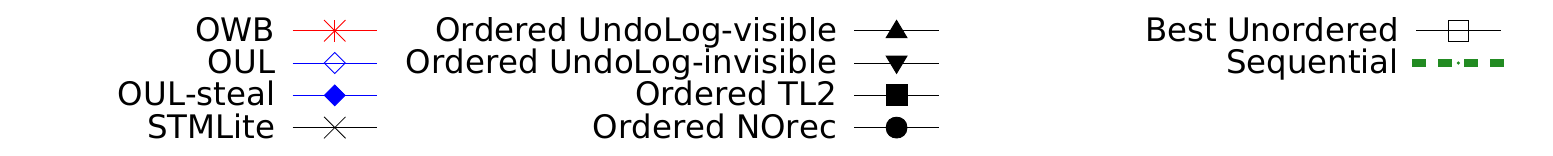}
\end{subfigure}
    \begin{subfigure}[b]{0.23\textwidth} 
    	\includegraphics[scale=0.35]{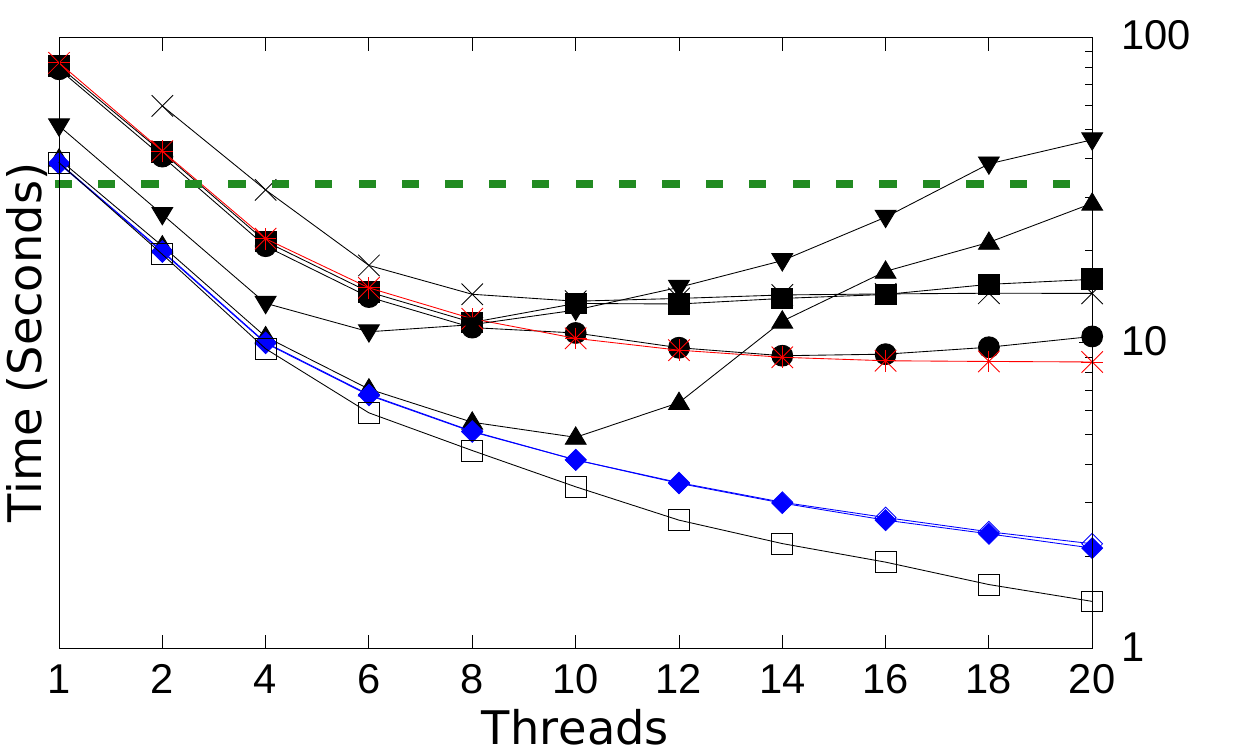} \vspace{-1.5em} \caption{Kmeans Low} \label{fig:kmeans-low} 
    \end{subfigure}
    \begin{subfigure}[b]{0.23\textwidth} 
    	\includegraphics[scale=0.35]{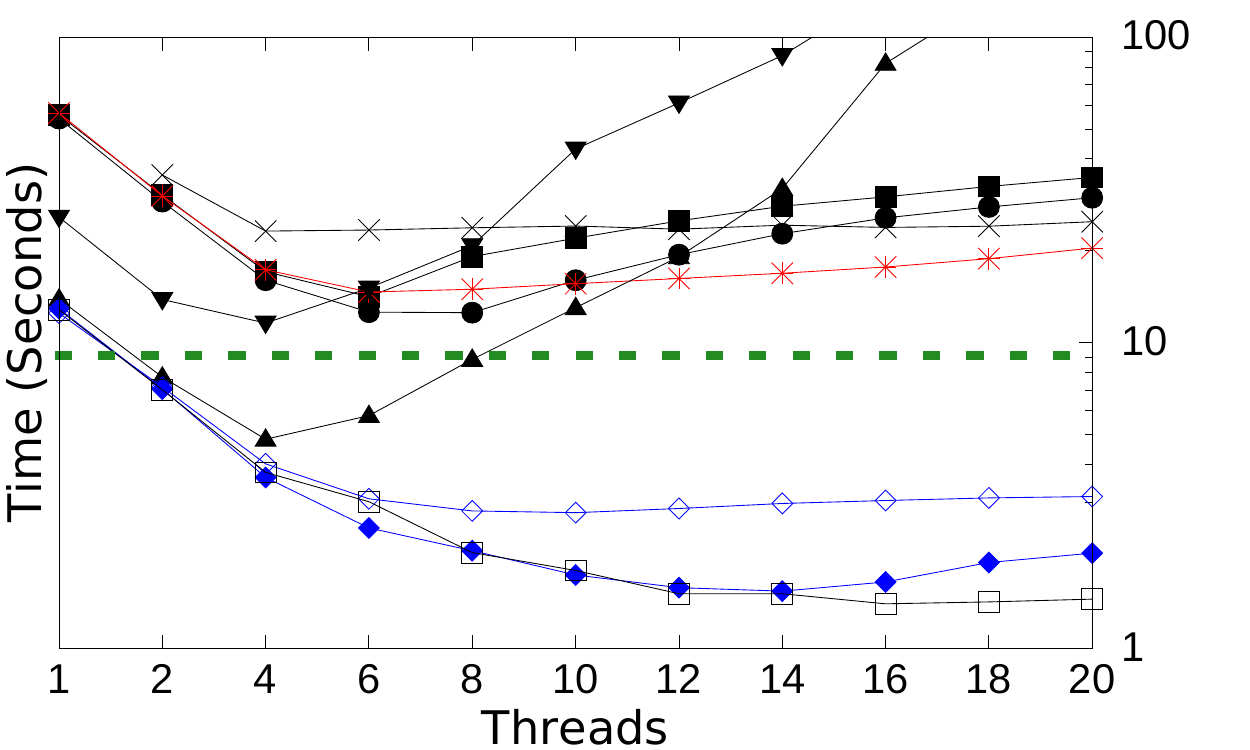} \vspace{-1.5em} \caption{Kmeans High} \label{fig:kmeans-high} 
    \end{subfigure}
    \begin{subfigure}[b]{0.23\textwidth} 
    	\includegraphics[scale=0.35]{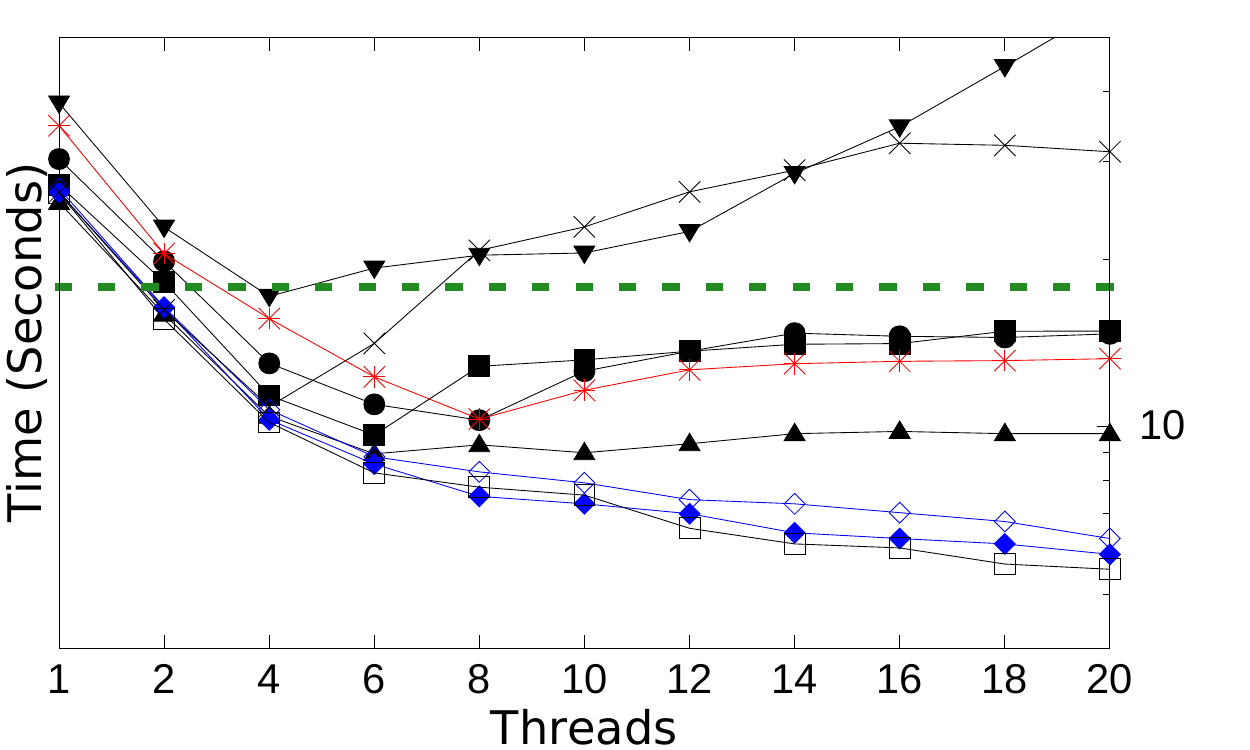} \vspace{-1.5em} \caption{Genome} \label{fig:genome} 
    \end{subfigure}
    \begin{subfigure}[b]{0.23\textwidth} 
    	\includegraphics[scale=0.35]{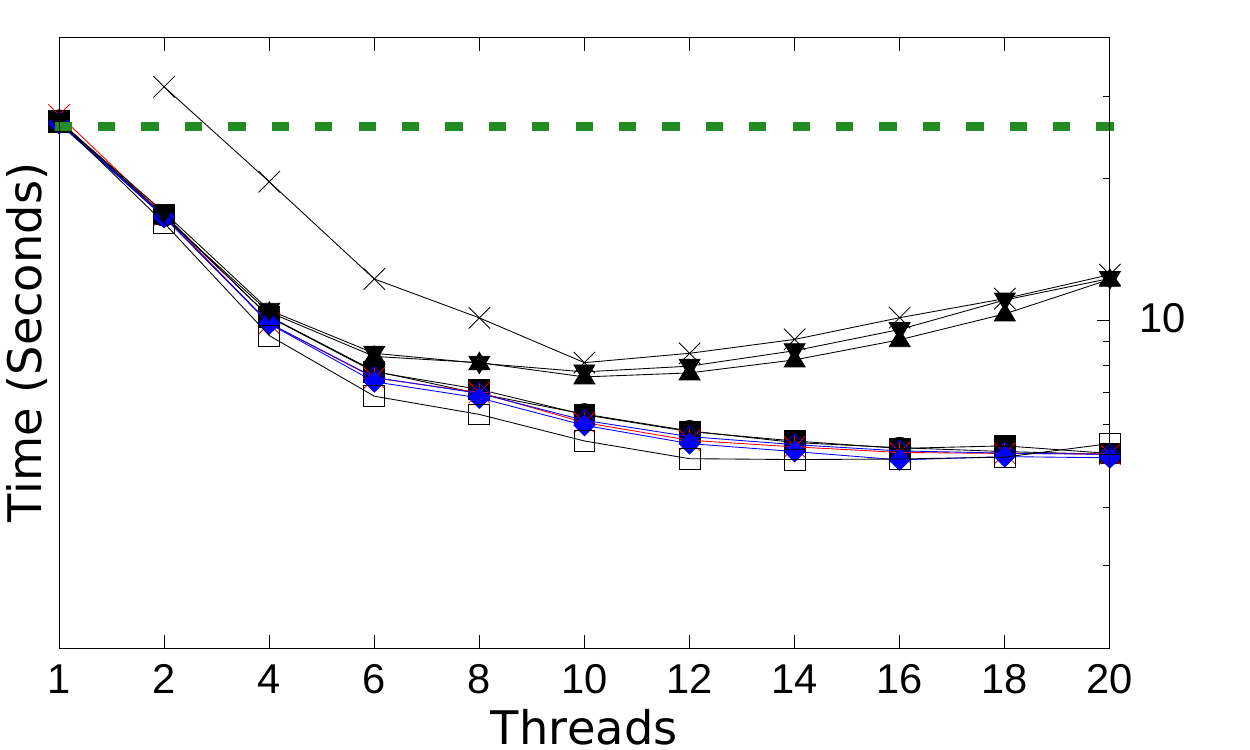} \vspace{-1.5em} \caption{SSCA2} \label{fig:ssca2} 
    \end{subfigure}
    \begin{subfigure}[b]{0.23\textwidth} 
    	\includegraphics[scale=0.35]{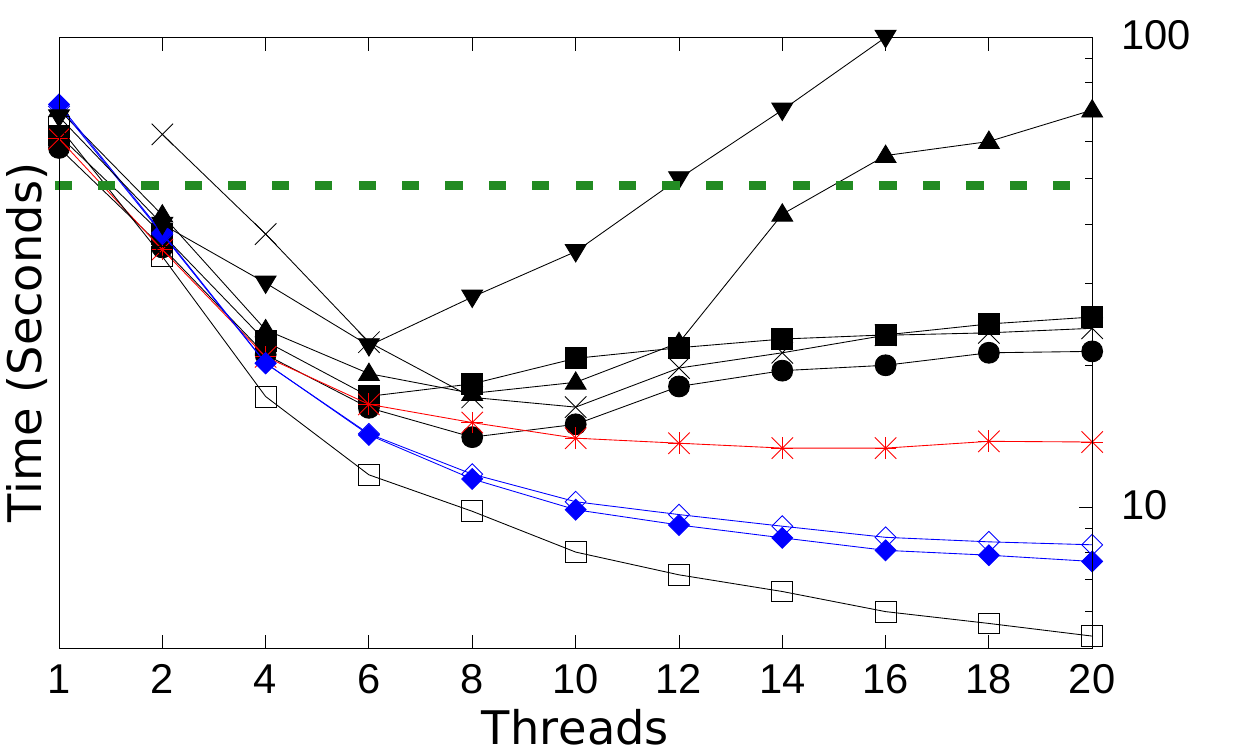} \vspace{-1.5em} \caption{Vacation Low} \label{fig:vacation-low} 
    \end{subfigure}
    \begin{subfigure}[b]{0.23\textwidth} 
    	\includegraphics[scale=0.35]{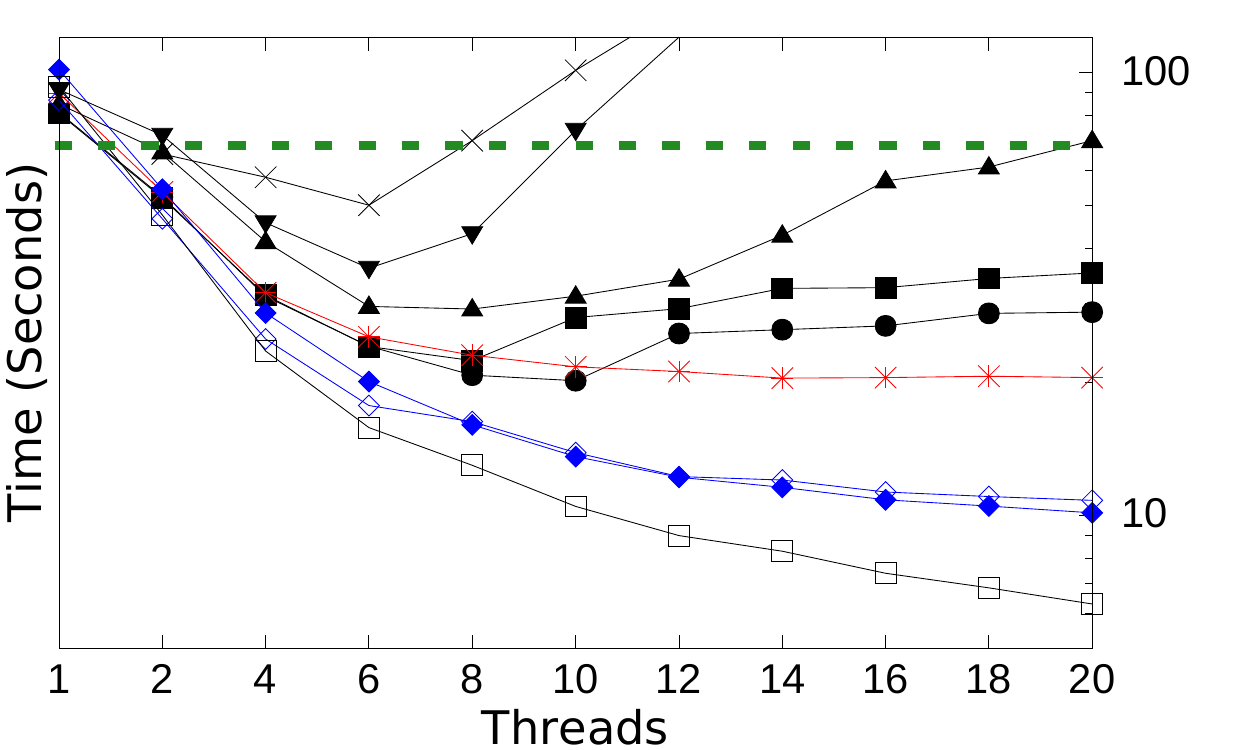} \vspace{-1.5em} \caption{Vacation High} \label{fig:vacation-high} 
    \end{subfigure}
    \begin{subfigure}[b]{0.23\textwidth} 
    	\includegraphics[scale=0.35]{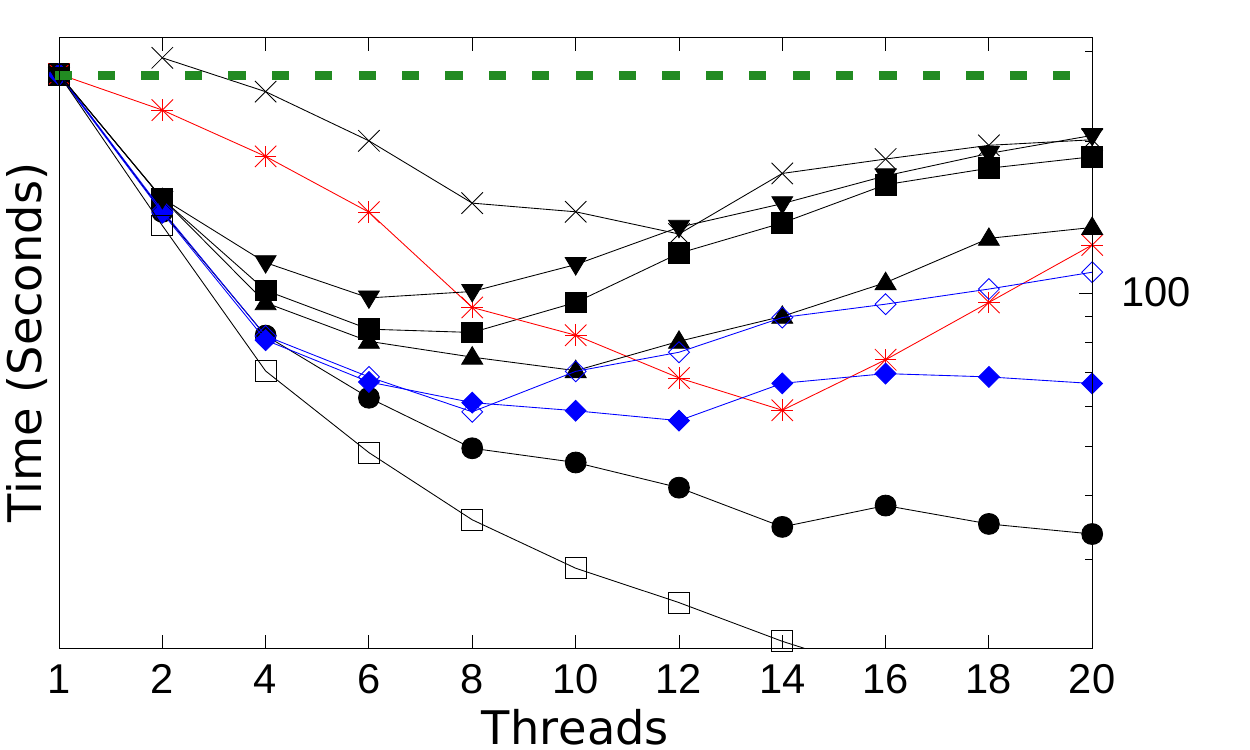} \vspace{-1.5em} \caption{Labyrinth} \label{fig:labyrinth} 
    \end{subfigure}
    \begin{subfigure}[b]{0.23\textwidth} 
    	\includegraphics[scale=0.35]{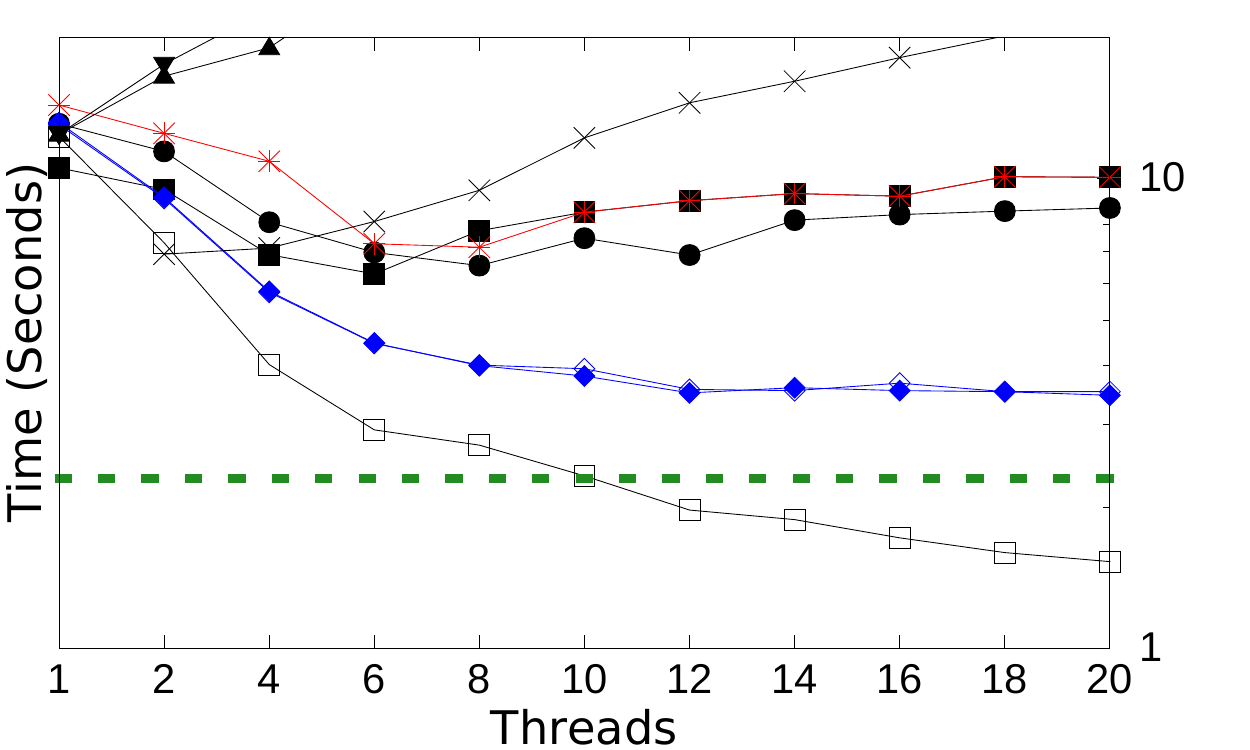} \vspace{-1.5em} \caption{Intruder} \label{fig:intruder} 
    \end{subfigure}
    \vspace{-0.5em}
    \caption{Execution time of STAMP (Y-axis log scale).}
    \label{fig:stamp}
%    \vspace{-15pt}
\end{figure}

\begin{figure*}
\centering
	\begin{subfigure}[b]{\textwidth} \centering \includegraphics[scale=0.3]{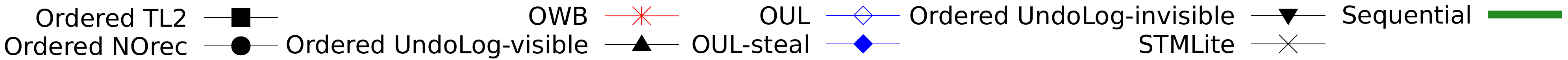}  \end{subfigure}
    \begin{subfigure}[b]{0.23\textwidth} \centering \includegraphics[scale=0.32]{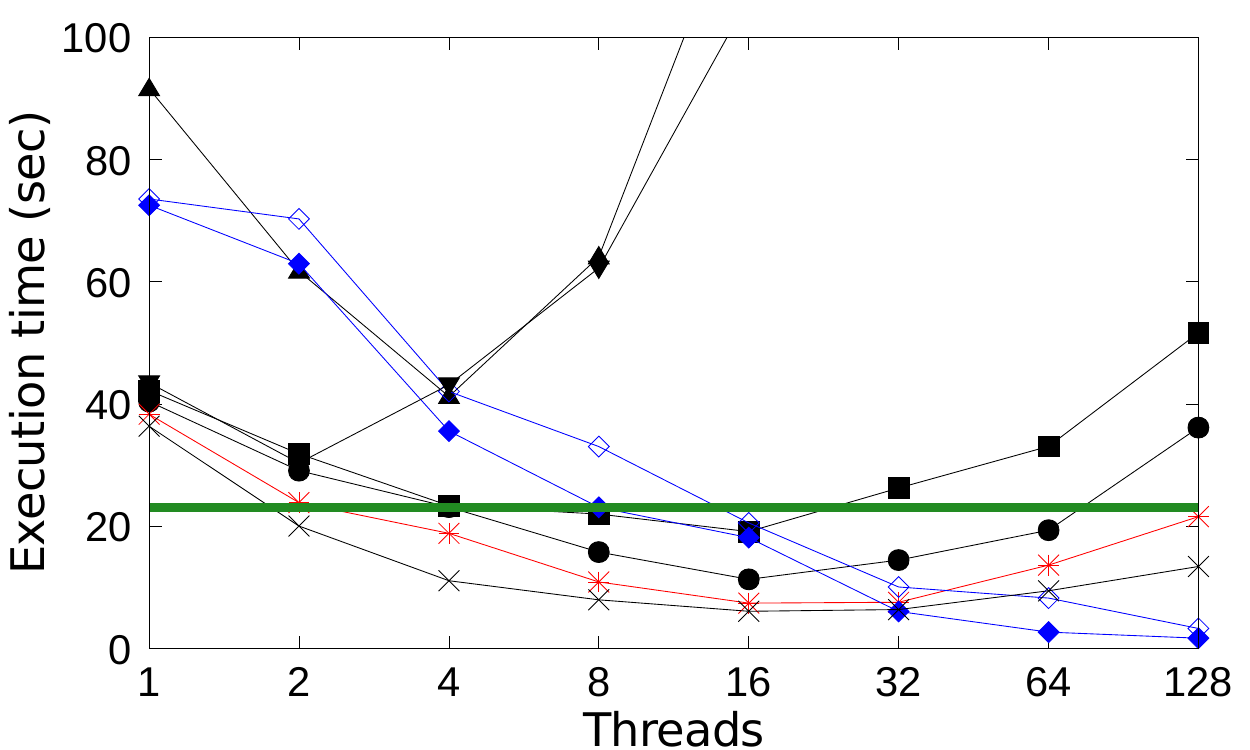}
    %\vspace{-1.7em}
    \caption{PARSEC/Blacksholes} \label{fig:blackshoels}
    \end{subfigure}    
    \begin{subfigure}[b]{0.23\textwidth} \centering \includegraphics[scale=0.32]{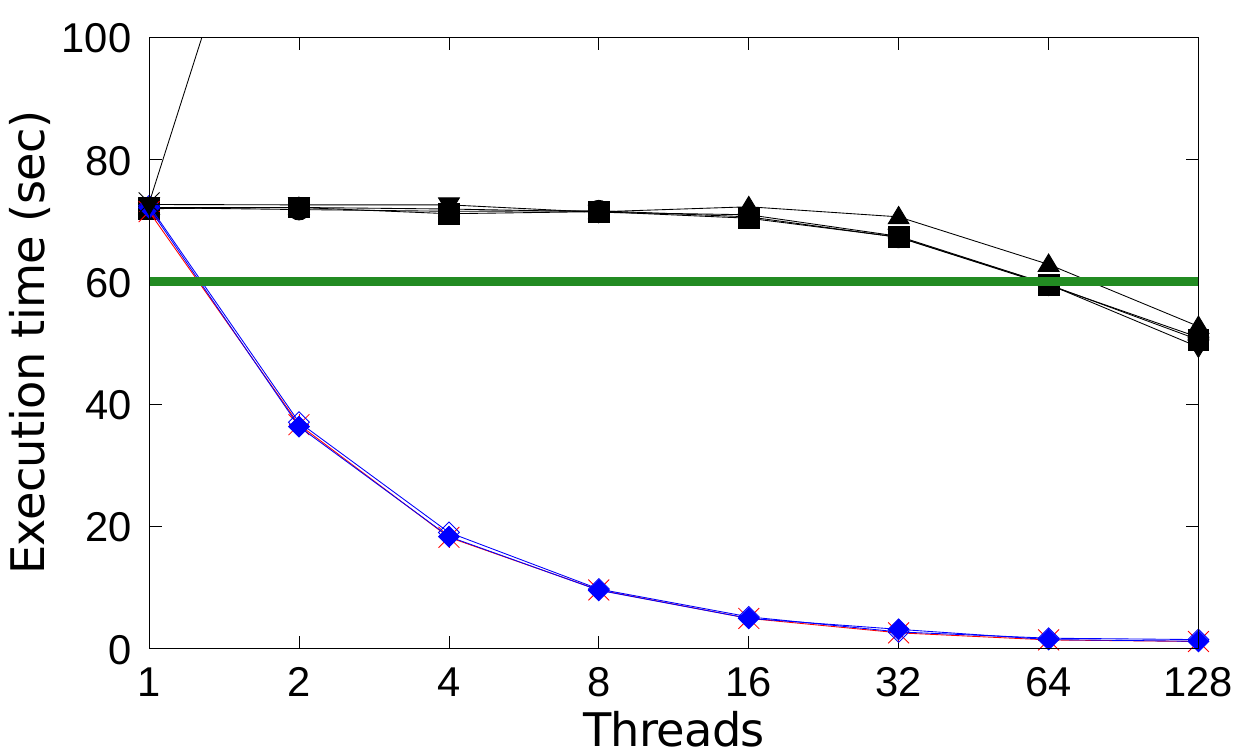}
%    \vspace{-1.7em}
    \caption{PARSEC/Swaptions}
    \label{fig:swamp}
    \end{subfigure}    
    \begin{subfigure}[b]{0.23\textwidth} \centering \includegraphics[scale=0.32]{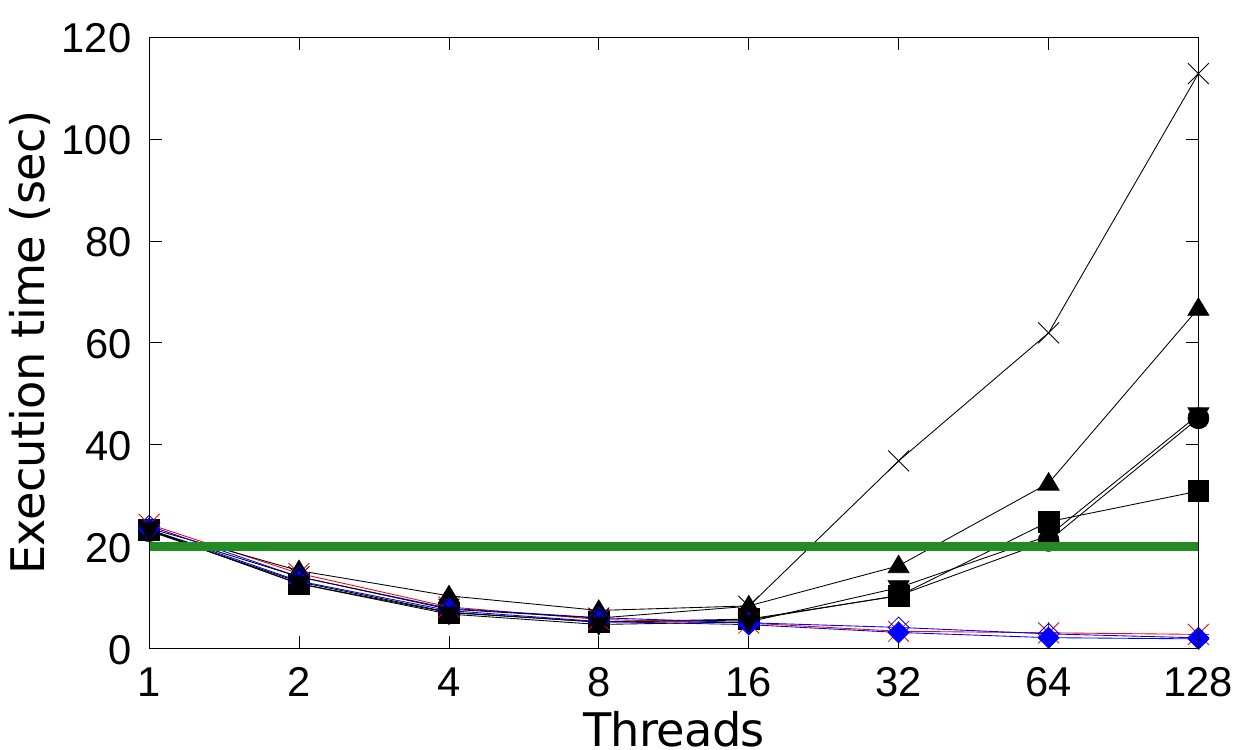}
    %\vspace{-1.7em}
    \caption{PARSEC/Fluidanimate}
    \label{fig:fluid}
    \end{subfigure}    
    \begin{subfigure}[b]{0.23\textwidth} \centering \includegraphics[scale=0.32]{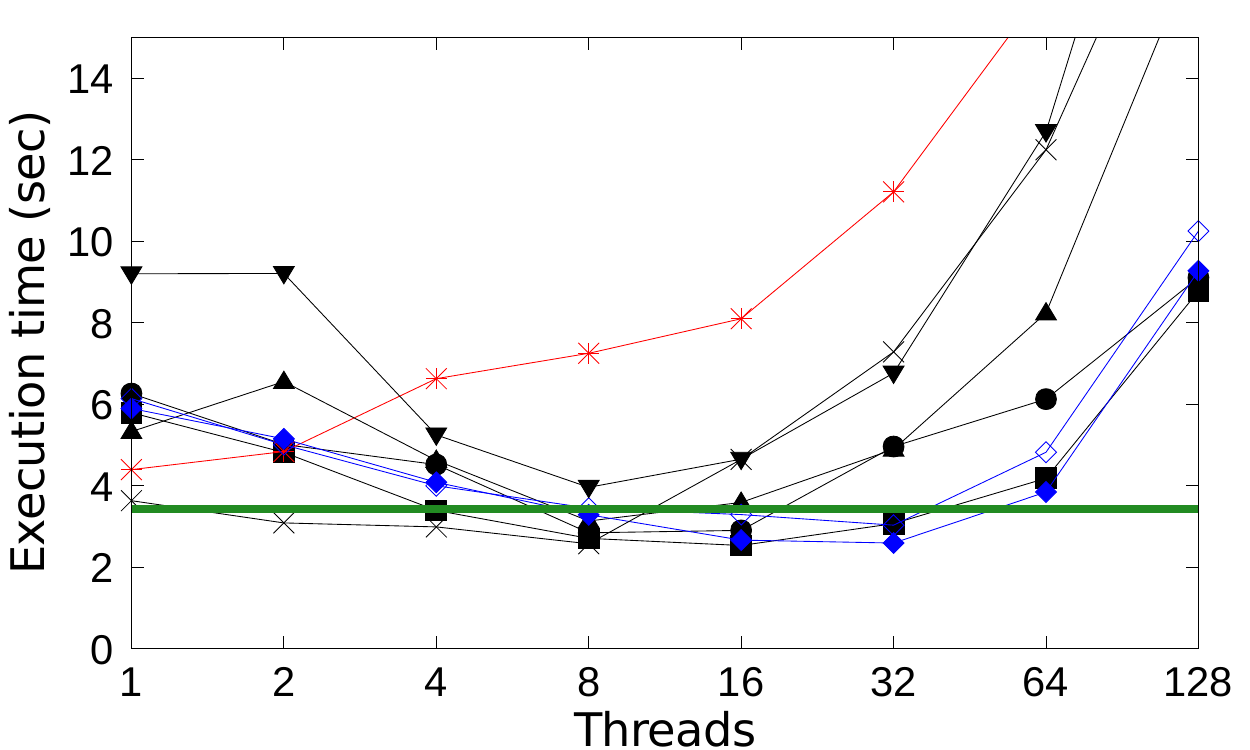}
    %\vspace{-1.7em}
    \caption{SPEC2000/Equake}
    \label{fig:equake}
    \end{subfigure}
%    \vspace{-2em}
    \caption{Execution time using PARSEC and SPEC200 benchmarks.} 
%    \vspace{-1.5em}
    \label{fig:spec}
\end{figure*}

\textbf{STAMP Benchmark.}
STAMP~\cite{caominh:stamp:iiswc:2008} is a benchmark suite with applications covering a variety of domains. 
Figure~\ref{fig:stamp} shows the execution time of the aforementioned algorithms (lower is better). Two applications (Yada and Bayes) have been excluded because they expose non-deterministic behaviors, thus their evolution is unpredictable. The datapoints for competitors that do not scale in some configuration are omitted to preserve the scale and readability of the plot. We also included performance of the unordered STM algorithm (among those in Figure~\ref{fig:micro-all}) that behaves best in each plot.

\textit{Kmeans}, a clustering algorithm, iterates over a set of points and associates them to clusters. The main computation is in finding the nearest point, while shared data updates occur when updating the cluster centers. 
Both OUL and OUL-Steal scale when increasing the number of workers, while under high contention OUL-Steal performs better (Figure~\ref{fig:kmeans-high}). OWB and Ordered NOrec have similar performance,
but OWB does not degrade at high thread count.

\textit{Genome} reconstructs the gene sequence from segments of a larger gene. It uses a shared hash-table to organize segments, which requires synchronization over its accesses. Genome exhibits a little contention, which makes OUL and OUL-Steal perform similarly (Figure~\ref{fig:genome}).

\textit{SCAA2} is a multi-graph kernel that is commonly used in domains such as biology and security. The core of the kernel uses a shared graph structure that is updated at each iteration. 
The amount of contention is low as the large number of graph nodes leads to infrequent concurrent updates. Figure~\ref{fig:ssca2} shows that all algorithms perform almost equally and benefit from optimistic concurrency.

\textit{Vacation} is a travel reservation system using an in-memory database. Each client uses a coarse-grain transaction to execute its session, consequently, aborts are costly. Again, our cooperative model boosts the performance of the proposed algorithms, and they scale well when increasing the number of workers (clients) (Figures~\ref{fig:vacation-low} and~\ref{fig:vacation-high}). 

\textit{Labyrinth} is a multi-path maze solver. The maze is represented as a three-dimensional uniform grid, and each thread tries to connect input pairs by a path of adjacent maze points. Upon finding a path, it is is highlighted at a shared output grid. Transactions conflict when their paths overlap. In Figure~\ref{fig:labyrinth}, NOrec outdoes other algorithms because of two reasons: \textit{1)} as Labyrinth updates adjacent addresses for the path, it is prone to produce false sharing for all other algorithms that use locks; and \textit{2)} NOrec employs a value-based validation, thus when two conflicting transactions updating a maze point with the same value, they commit successfully.

\textit{Intruder}, a network intrusion detection system using signatures. It compares the captured packets against a dictionary of intrusion signatures. Packets are processed in parallel, grouped in sessions, and stored in a self-balanced (red-black) tree. Transactions guard the tree operations, and the contention is high and depends on the frequency of the rebalance operation. Figure~\ref{fig:intruder} shows that not all algorithms scale well; besides, the sequential execution outperforms all of them (except the unordered).

\textbf{PARSEC Benchmark.}
PARSEC is a benchmark suite for shared memory chip-multiprocessors architectures.

The \textit{Black-Scholes} application calculates Black-Scholes equation for input values. Since calculations per iteration are few, each transaction involves multiple calculations to reduce the overhead or parallelization.
\textit{Swaptions} employs Monte Carlo simulation to compute prices. \textit{Fluidanimate} is an application performing physics simulations. The main computation is spent on computing particle densities and forces, which involves six levels of loops nesting updating a shared array structure.  Since it is not straightforward to assign ages based on number of iterations, a global atomic integer variable is used to assign ages to transactions.

OUL, OUL-Steal and OWB scale in these three applications; significant speedup over sequential is achieved in Swaptions.
In both Black-Scholes and Fluidanimate, all other algorithms outperform sequential when contention is low, and then performance drop quickly when contention increases, which is due to a high abort rate.

\textbf{SPEC CPU2000 Benchmark.}
\textit{Equake} is an application included in the SPEC CPU2000 benchmark and it simulates the propagation of elastic waves. The computation iterates over a number of steps and, in each time step, it iterates over a number of nodes where each performed calculation relies on the previous one. The loop-carried dependencies forces the transaction to be committed in a specific order. Each thread is assigned a consecutive region of nodes so only those in joints may abort. 

When testing this benchmark, we set the input size to be 500 nodes. The results show that OUL, OUL-Steal and OTL2 scales when increasing the number of threads, up to 32 threads; the achieved peak speedup is 30\%. After that, because of high contention and increasing number of aborts, all algorithms' performance drops.
%OUL-Steal can write through shared object by stealing locks thus write-write conflicts have less influence on performance.

\section{Conclusion}
In this paper, we presented three algorithms that address the problem of committing transactions with an order defined prior to execution. Our results show that even if a system requires a specific commit order, it is possible to achieve high performance exploiting parallelism with data conflicts.

\section*{Acknowledgments}
Authors would like to thank the anonymous reviewers for their valuable comments, Binoy Ravindran for his feedback at the very early stage of this paper, and Jacob Nelson for the insightful discussion. This material is based upon work supported by the Air Force Office of Scientific Research under award number FA9550-17-1-0367 and by the National Science Foundation under Grant No. CNS-1814974.

\bibliographystyle{ACM-Reference-Format}
\bibliography{all}
\end{document}